\documentclass[referee,letterpaper]{aa}

\usepackage{txfonts}
\usepackage{graphicx}
\usepackage{natbib}
\usepackage{supertabular}

\begin{document}

\title{A new paradigm for the X-ray emission of O stars from {\it XMM-Newton}
       observations of the O9.7 supergiant \hbox{$\zeta$~Orionis}
       \thanks{Based on observations obtained with {\it XMM-Newton}, an ESA science mission with instruments
               and contributions directly funded by ESA Member States and NASA}
}
\titlerunning{A new O-star X-ray paradigm}

\author{A. M. T. Pollock}
\authorrunning{Pollock}

\institute{European Space Agency {\it XMM-Newton} Science Operations Centre,
           European Space Astronomy Centre,
           Apartado 50727,
           Villafranca del Castillo,
           28080 Madrid,
           Spain.}

\date{Received 2005 July 15 / Accepted 2006 November 09}

\abstract{{\it XMM-Newton} observations of the O supergiant \hbox{$\zeta$~Orionis} (O9.7 Ib)
          extend knowledge of its high-resolution spectrum beyond the \ion{C}{vi} line at 33.7\AA\
          and suggest a new framework for the interpretation of the X-ray spectra of single hot stars.
          All the lines are broad and asymmetric with similar velocity profiles.
          X-rays probably originate in the wind's terminal velocity regime in collisionless shocks
          controlled by magnetic fields rather than in cooling shocks
          in the acceleration zone. During post-shock relaxation,
          exchange of energy between ions and electrons is so slow that electron
          heating does not take place before hot gas is quenched by the majority cool gas.
          The observed plasma is not in equilibrium and the electron bremsstrahlung
          continuum is weak. Charge exchange, ionization and excitation are likely to be produced
          by protons. Fully thermalized post-shock velocities ensure high cross-sections
          and account for the observed line widths, with some allowance probably necessary
          for non-thermal particle acceleration.
          In general, the form of X-ray spectra in both single and binary stars is likely
          to be determined principally by the amount of post-shock electron heating:
          magnetically confined X-ray plasma in binary systems can evolve further
          towards the higher electron temperatures of equilibrium
          while in single stars this does not take place.
          The long mean-free path for Coulomb energy exchange between fast-moving ions
          may also inhibit the development of line-driven instabilities.
}

\maketitle

\section{Introduction\label{Section:Introduction}}

The O9.7~Ib supergiant \object{\hbox{$\zeta$~Orionis}}, or Alnitak, is the
easternmost star of Orion's belt and lies just above the Horsehead
Nebula. In addition to its glorious appearance in the night sky, it
is a prominent star in other respects. It is optically the brightest
O star and, at X-ray frequencies, it is the seventh most luminous in
absolute terms and the fourth in apparent brightness
\citep{BSC:1996}. \cite{WC:2001} observed it early with the {\it
Chandra} high-energy gratings to yield one of the first published
high-resolution X-ray spectra and argued for the origin of at least
some of the X-rays very close to the stellar surface at the base of
the powerful wind that is a ubiquitous feature of such hot stars.
The intensity ratios of the He-like triplets, which are susceptible
to modification by the strong photospheric UV radiation field, were
used to locate the site of production.

Its X-ray spectrum has proved
to be typical of those seen from O stars with very prominent broad
lines from a variety of H-like and He-like ions of common light
elements and L-shell lines from \ion{Fe}{xvii} in particular.
\cite{MCWMC:2002} and \cite{M:2002} commented on the unexpected and little understood line
widths observed from star to star and made general efforts to
reconcile the data with the popular view that shocks developing from
instabilities in the wind line-driving mechanism are responsible for
generating the X-rays. Whatever their origin, \cite{FKPPP:1997}
suggested that X-ray temperature shocks cool very rapidly via
radiation losses in the inner wind, probably accounting for the
range of temperatures of one to a few million degrees observed in
even comparatively low-resolution ROSAT PSPC spectra. It is clear
from the high-resolution spectra that a range of apparent
temperatures is
implied by the simultaneous presence of prominent lines such as
\ion{O}{vii} typical of 0.8MK gas; \ion{O}{viii} of 2.4MK; 
\ion{Fe}{xvii} of 4MK; and \ion{Si}{xiii} of 10MK.

{\it XMM-Newton} is ideal for observations of such hot stars. The bandwidth of the Reflection Grating
Spectrometer (RGS) matches perfectly that part of the X-ray spectrum in which the lines
occur, extending the {\it Chandra}  High-Energy Transmission Grating (HETG)
spectra beyond 25\AA\ to the \ion{N}{vi} lines near 29\AA\
and \ion{C}{vi} near 34\AA\ into the temperature regime below a million degrees where the
radiative emissivity increases rapidly. The RGS is able to resolve
the lines and its high sensitivity allows the accumulation of enough statistics
to study line profiles in detail.

The supergiant in which we are interested is the dominant component of a multiple system
\citep{HWEHN:2000} and is more properly known as \hbox{$\zeta$~Orionis~Aa},
though we refer to it simply as \hbox{$\zeta$~Orionis} throughout.
Its fainter companions are \hbox{$\zeta$~Orionis~Ab} at 0.045\arcsec and
\hbox{$\zeta$~Orionis~B} at 2.4\arcsec, neither of which is likely to contribute to the X-ray emission;
and \hbox{$\zeta$~Orionis~C} at 58\arcsec, which would be easily resolvable by all the{\it XMM} instruments.
\hbox{$\zeta$~Orionis~B} was resolved in the {\it Chandra} observation at an intensity
of a few percent of \hbox{$\zeta$~Orionis}.
Some relevant parameters which we have adopted for \hbox{$\zeta$~Orionis} are listed in Table~\ref{Table:zO:data}.
Because the terminal velocity is an important parameter, it is worth bearing in mind
that earlier work by \cite{PBH:1990} preferred a value of $1860~{\rm km}~{\rm s}^{-1}$,
the difference probably reflecting the uncertainty in such measurements.

\begin{table*}[ht]
\caption{\label{Table:zO:data}
         Relevant data for HD37742 O9.7Ib \hbox{$\zeta$~Orionis} from \cite{LHdKL:1999} unless stated otherwise.}
\begin{center}
\begin{tabular}{lccl}
\hline
visual magnitude            & $V$           &  2.03                                             &                     \\
effective temperature       & $T_{\rm eff}$ &  $30900~{\rm K}$                                  &                     \\
radius                      & $R_*$         &  $31~{\rm R}_{\odot}$                             &                     \\
terminal velocity           & $v_{\infty}$  &  $2100~{\rm km}~{\rm s}^{-1}$                     &                     \\
mass-loss rate              & $\dot{M}$     &  $1.4\times10^{-6}~{\rm M}_{\odot}~{\rm yr}^{-1}$ &                     \\
distance                    & $d$           &   473~pc                                          & \cite{dZ:etal:1999} \\
interstellar column density & $N_{\rm H}$   &  $2.5\times10^{20}~{\rm cm}^{-2}$                 & \cite{DS:1994}      \\
\hline
\end{tabular}
\end{center}
\end{table*}

\section{Observations\label{Section:Observations}}

The European Space Agency's {\it XMM-Newton} Observatory \citep{J:etal:2001} has been
in operation since 1999. Some of its operational and instrumental
characteristics and a range of initial astrophysical results were
described in the special 2001-01-01 issue of Astronomy \& Astrophysics,
including the high-resolution RGS spectrometers by \cite{dH:etal:2001}
and the EPIC imaging moderate-resolution spectrometers by
\cite{S:etal:2001} and \cite{T:etal:2001}.
{\it XMM} was used to observe \hbox{$\zeta$~Orionis} for nearly
12 hours in September 2002 as part of
guaranteed time. All the instruments operate simultaneously except, in this case,
the Optical Monitor (OM) \citep{M:etal:2001}, which was not used
because of the extreme optical brightness of the target.
Some observational details are shown in Table~\ref{Table:XMM:log}.
The data were reduced with standard procedures using the
{\it XMM} Science Analysis System SASv6.5.0 with
the calibration data available up to July 2005.

\begin{figure*}
\begin{center}
\rotatebox{90}{\scalebox{0.55}{\includegraphics{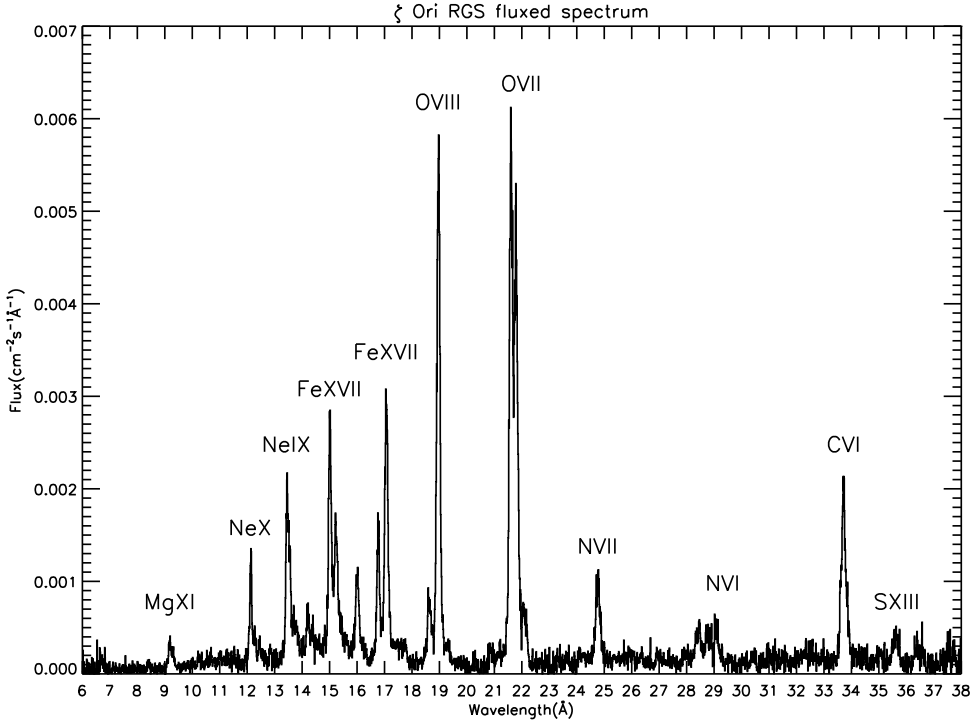}}}
\end{center}
\caption{\label{Figure:RGS:spectrum}
    The RGS spectrum of \hbox{$\zeta$~Orionis} on 2002-09-15. The spectrum is a combination of all
    the available RGS1 and  RGS2 1st and 2nd order data
    corrected for instrumental sensitive area using the SAS task {\tt rgsfluxer}. }
\end{figure*}

The combined RGS spectrum in Fig.~\ref{Figure:RGS:spectrum}
shows that the type of strong lines seen
in the HETG spectra below 25\AA\ continue to the longest RGS wavelengths.
The He-like \ion{O}{vii} triplet was the strongest line followed by \ion{O}{viii} Ly$\alpha$;
the \ion{Fe}{XVII} lines between 15 and 17~\AA;
and the strong detection of \ion{C}{vi} Ly$\alpha$ near 34\AA.
Where they
overlap between about 6 and 25~\AA, the RGS and HETG spectra are essentially identical as
discussed in more detail below.

\begin{figure*}
\begin{center}
\rotatebox{270}{\scalebox{0.55}{\includegraphics{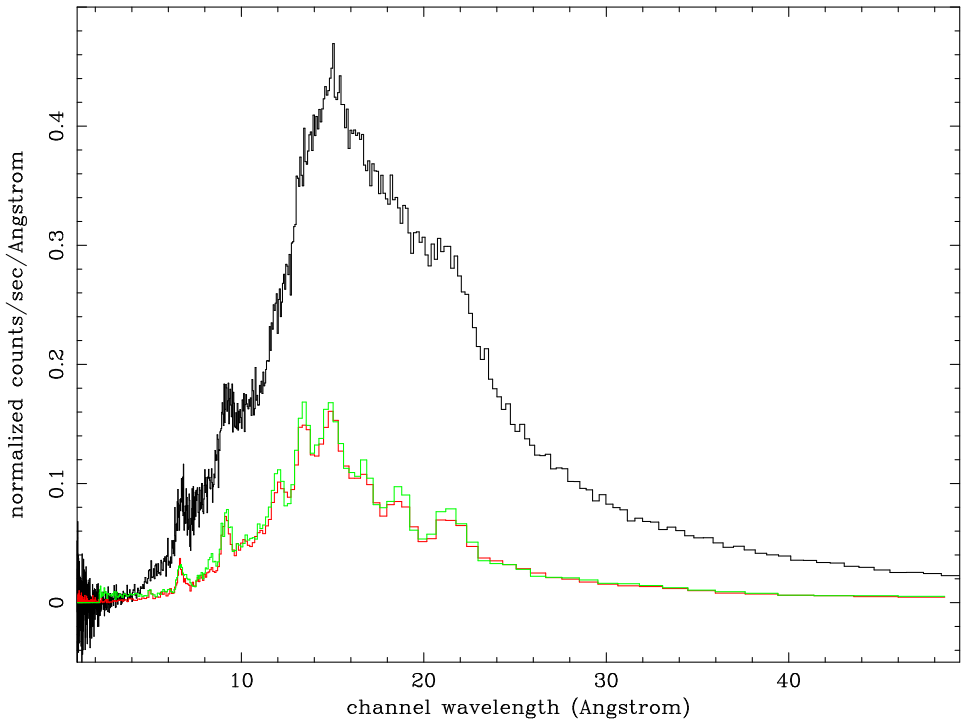}}}
\end{center}
\caption{\label{Figure:EPIC:spectra}
    Spectra of \hbox{$\zeta$~Orionis} on 2002-09-15 from the three EPIC instruments. 
    The pn, shown in black, had the highest count rate
    mainly because it does not share its telescope module with an RGS. The MOS instruments have better
    energy resolution and were able to separate lines reasonably well. 
    The agreement was good between MOS1 and MOS2
    which were operated in different modes. MOS1, shown in red, was in
    small-window imaging mode while MOS2 was in timing mode.
    }
\end{figure*}

The EPIC spectra are shown in
slightly different form in Fig.~\ref{Figure:EPIC:spectra} without
any effective area corrections as this becomes misleading when the
energy resolution is relatively poor. Nonetheless, the EPIC data
complement the RGS because of their higher sensitivities, as
emphasized by the detected count rates in Table~\ref{Table:XMM:log},
and wider bandwidths. Both pn and MOS show that almost all
\hbox{$\zeta$~Orionis}'s X-rays fall in the RGS waveband. At short
wavelengths beyond the strong MOS detections of the He-like triplets
\ion{Mg}{xi} and \ion{Si}{xiii}, apart from weak detections of
\ion{Si}{xiv} and \ion{S}{xv}, there is no evidence of a hard X-ray
tail. At the longest wavelengths, the counts are overwhelmingly due
to redistribution from higher energies as the CCD energy resolutions
deteriorate due to surface effects. Despite the limited ability to
distinguish individual lines, the high pn count rate provides the
most effective means of assessing source variability of which there
was none detectable during this observation. \cite{BS:1994} made a
variability study of \hbox{$\zeta$~Orionis} with {\it ROSAT} and
found effectively no evidence for either short-term or long-term
variability although there was one isolated episode during which the
X-ray count rate increased by $15\%$.

\begin{table*}[ht]
\caption{\label{Table:XMM:log}
         {\it XMM-Newton} \hbox{$\zeta$~Orionis} observation log showing the background-corrected count rates. The RGS values
         are the 1st order rates.}
\begin{center}
\begin{tabular}{lll@{\ }cccr@{$\pm$}l}
\hline
Instrument    & Filter
              & Mode
              & \multicolumn{2}{c}{Start and End Dates}
              & Duration(s)
              & \multicolumn{2}{c}{Count Rate(${\rm s}^{-1}$)} \\
\hline
RGS1          &  ...    & Spec+Q       & 2002-09-15T13:11:17 & 2002-09-16T00:52:38 & 41979 & 0.346 & 0.003 \\
RGS2          &  ...    & Spec+Q       & 2002-09-15T13:11:17 & 2002-09-16T00:52:38 & 41979 & 0.322 & 0.003 \\
MOS1          &  Thick  & Small Window & 2002-09-15T13:12:32 & 2002-09-16T00:51:00 & 41728 & 1.735 & 0.007 \\
MOS2          &  Thick  & Timing       & 2002-09-15T13:12:28 & 2002-09-16T00:46:41 & 41473 & 1.692 & 0.006 \\
pn            &  Thick  & Full Frame   & 2002-09-15T14:04:08 & 2002-09-16T00:51:20 & 38362 & 6.659 & 0.014 \\
\hline
\end{tabular}
\end{center}
\end{table*}

\section{Shape and strength of X-ray lines in \hbox{$\zeta$~Orionis} \label{Section:Lines}}

\begin{figure}
\begin{center}
\rotatebox{90}{\scalebox{0.3}{\includegraphics{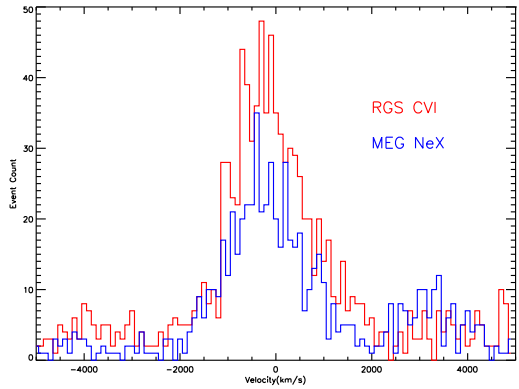}}} \\
\end{center}
\caption{\label{Figure:RGS:CVI:MEG:NeX}
    The RGS \ion{C}{vi} and MEG \ion{Ne}{x} Ly$\alpha$ velocity profiles
    of \hbox{$\zeta$~Orionis}. The resolutions of the two instruments are roughly
    the same in velocity units at these wavelengths, allowing direct comparison of the two
    profiles. The MEG feature at about $+3000~{\rm km}~{\rm s}^{-1}$ is due to
    \ion{Fe}{xvii} and \ion{Fe}{xxi} .}
\end{figure}

\begin{figure}
\begin{center}
\rotatebox{90}{\scalebox{0.3}{\includegraphics{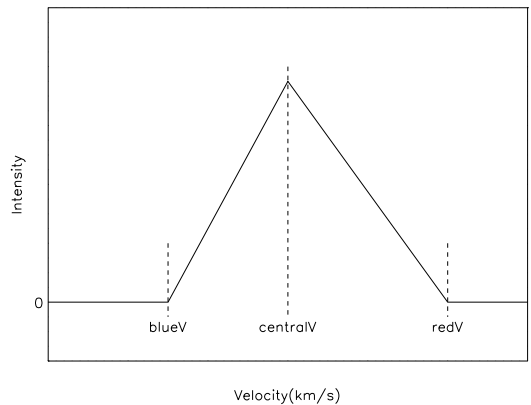}}} \\
\end{center}
\caption{\label{Figure:TriLine:model}
    The {\tt TriLine} model incorporated into {\tt XSPEC} to describe the X-ray lines
    of \hbox{$\zeta$~Orionis}. A similar {\tt SkewLine} model with unequal red and blue
    Gaussian widths was also used.}
\end{figure}

Comparison, for example, of the Ly$\alpha$~lines of
\ion{C}{vi}~$\lambda33.734$ and \ion{Ne}{x}~$\lambda12.132$
separated by nearly a factor of 3 in wavelength shows they had
shapes in velocity space that were very similar as shown in
Fig.~\ref{Figure:RGS:CVI:MEG:NeX}.
This similarity appears to apply to all the lines
and we have
been able to synthesize a good fit to the entire observed high-resolution spectrum
with the same velocity profile for every line. In order to model
the type of line asymmetries in \hbox{$\zeta$~Puppis}
discovered by \cite{KLC:etal:2001} and \cite{CMWMC:2001},
we added to {\tt XSPEC}\footnote{http://xspec.gsfc.nasa.gov/docs/xanadu/xspec/index.html}
a local triangular line-profile model named {\tt TriLine}
characterized,
as illustrated in Fig.~\ref{Figure:TriLine:model},
by three shape parameters:
independent red and blue velocities where the profile goes to zero
and a velocity shift of the central peak from the laboratory wavelength.

The model included 85 lines from 18 ions with {\tt TriLine} profiles,
with laboratory wavelengths taken from the
APED \citep{SBLR:2001} and CHIANTI \citep{Y:etal:2003}
line lists,
and a thermal continuum subject to interstellar
absorption fixed at the value given in Table~\ref{Table:zO:data}.
The 18 model ions were the H-like ions of 
\ion{C}{vi}, \ion{N}{vii}, \ion{O}{viii}, \ion{Ne}{x}, \ion{Mg}{xii} and \ion{Si}{xiv};
the He-like ions of
\ion{N}{vi}, \ion{O}{vii}, \ion{Ne}{ix}, \ion{Mg}{xi} and \ion{Si}{xiii};
\ion{Fe}{xvii-xxi}; and \ion{S}{xii-xiii}.
The Lyman series lines were explicitly treated as doublets with the
flux ratio fixed at 2:1.
The continuum was
intended to account empirically for the combination of
true electron continuum and other lines.
The nature of the continuum is discussed further below in Section~\ref{Section:Continuum}.

Fits were not made to fluxed spectra like that plotted in
Fig.~\ref{Figure:RGS:spectrum} but to the detected counts spectra
using the appropriate response matrices. Many spectral bins
contain few or even zero counts. For such photon-limited data
to which Poissonian rather than Gaussian statistics apply,
a likelihood measure like that provided by {\tt XSPEC}'s C-statistic
is the correct choice rather than a $\chi^{2}$-statistic.
Errors reported in this paper were calculated for \hbox{$\Delta{\rm C}=1$}.

The best-fit parameters reported in Tables~\ref{Table:TriLine:model} and
\ref{Table:TriLine:normalisations} were calculated for a combined
fit to the spectra of RGS1 and RGS2 1st and 2nd order and the {\it Chandra}
MEG and HEG $\pm1$ spectra retrieved from the public archive.

These models were a significant technical challenge with the large
number of different lines and spectra involved, to accommodate
which an expanded version of {\tt XSPEC v11}
was built. There were 3472 parameters in the model
which reduced to 84 for the single line profile
once all the constraints were applied through {\tt XSPEC}'s Tcl script capabilities.

\begin{figure}
\begin{center}
\begin{tabular}{c}
\rotatebox{90}{\scalebox{0.25}{\includegraphics{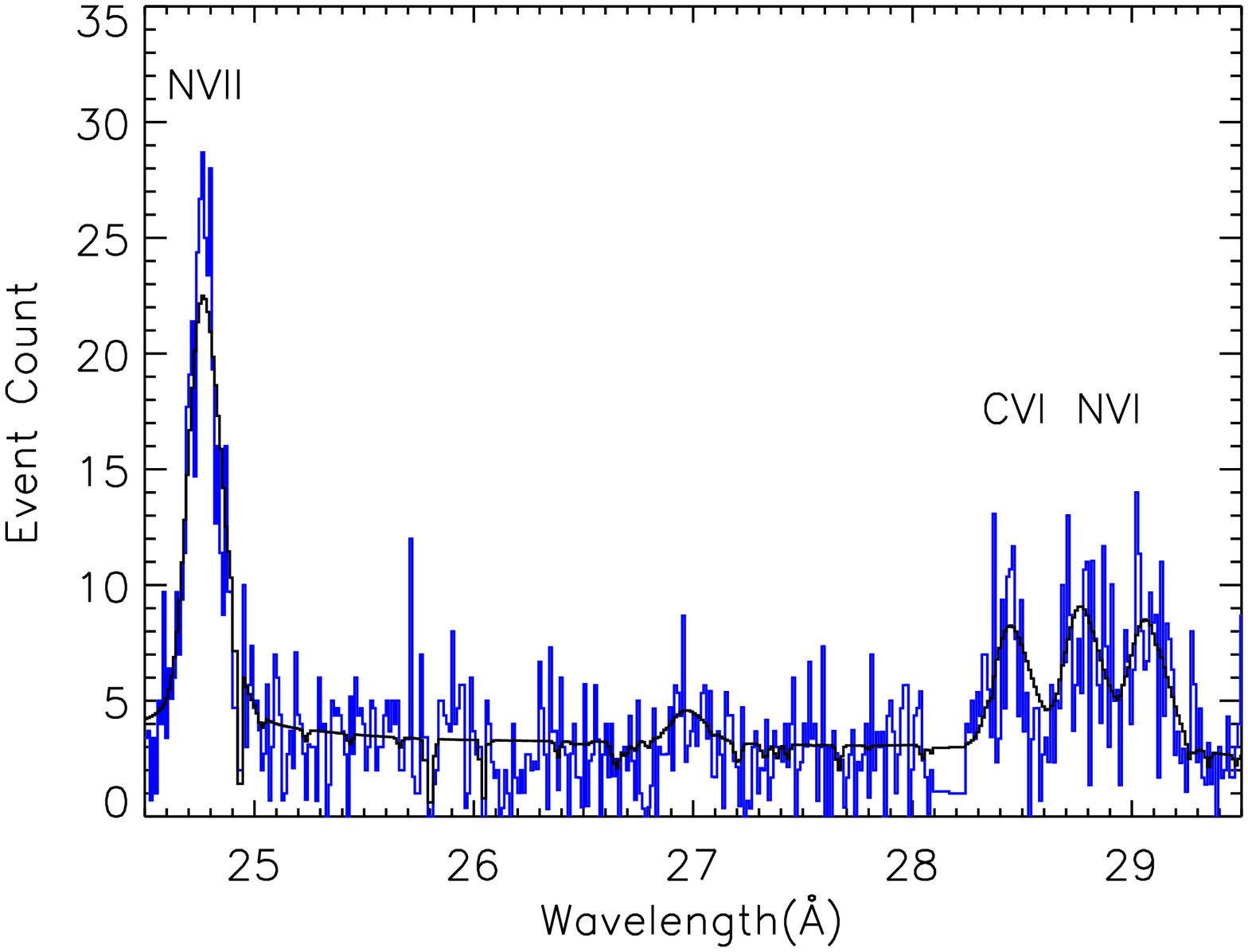}}} \\
\rotatebox{90}{\scalebox{0.25}{\includegraphics{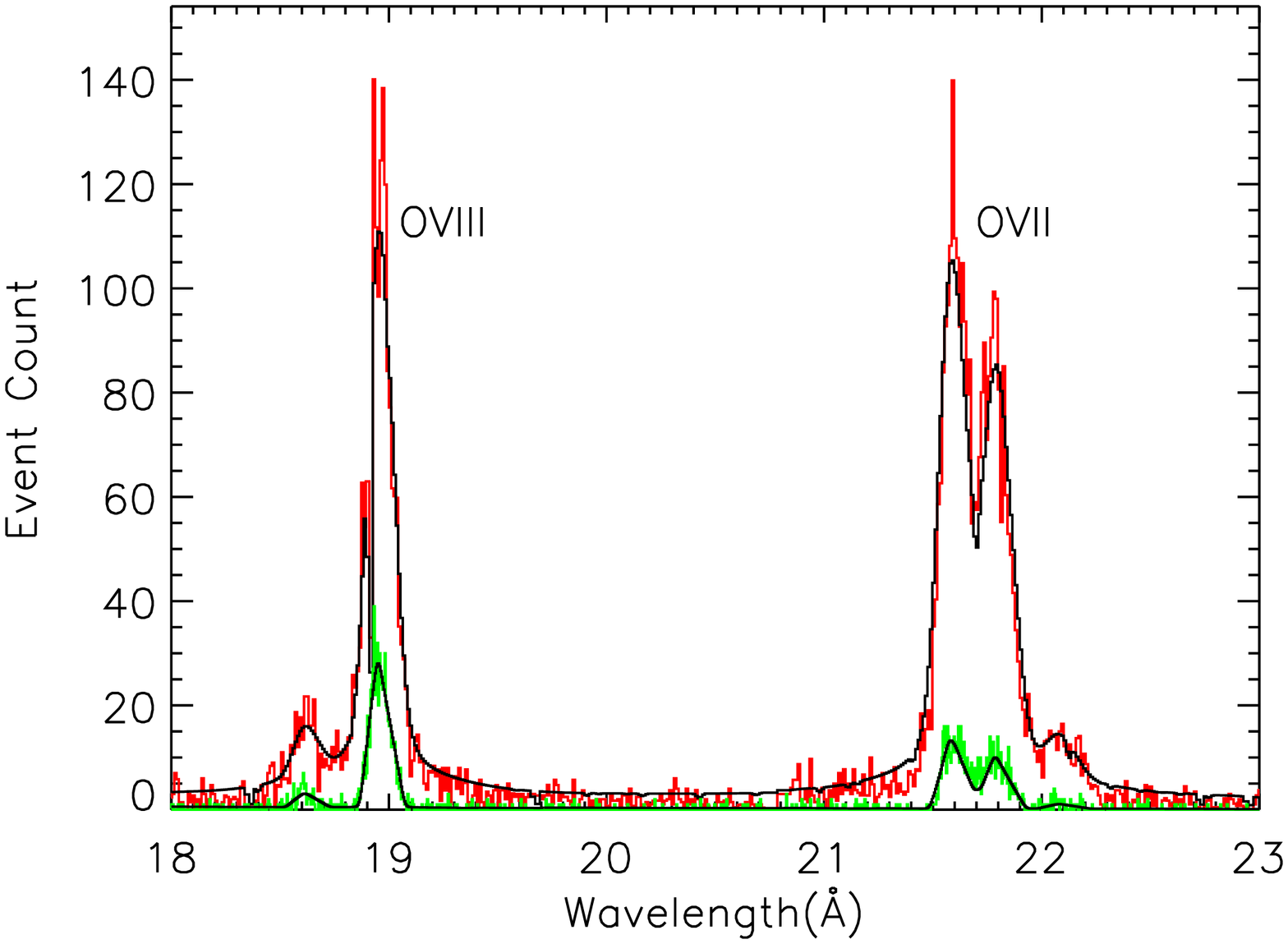}}} \\
\rotatebox{90}{\scalebox{0.25}{\includegraphics{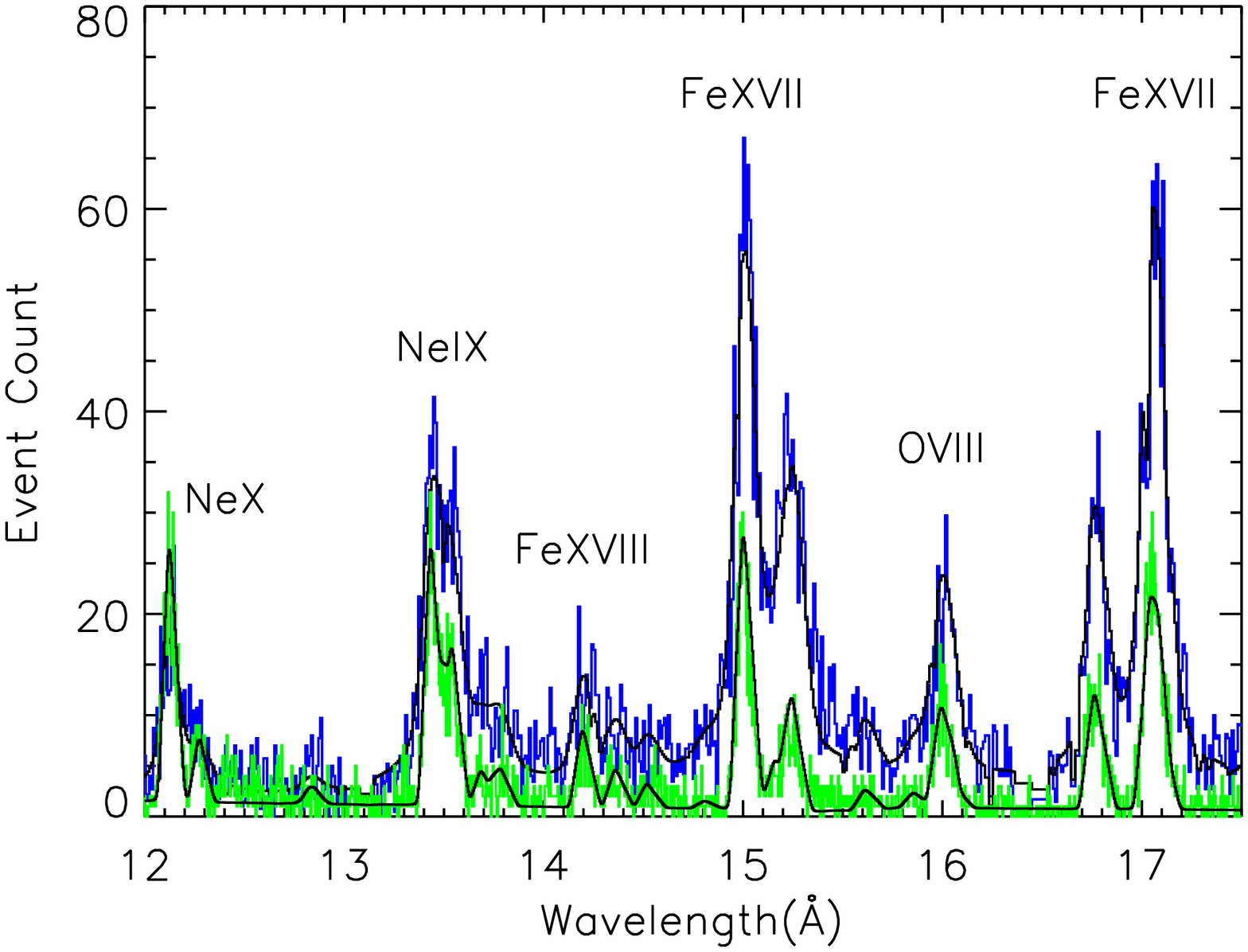}}} \\
\rotatebox{90}{\scalebox{0.25}{\includegraphics{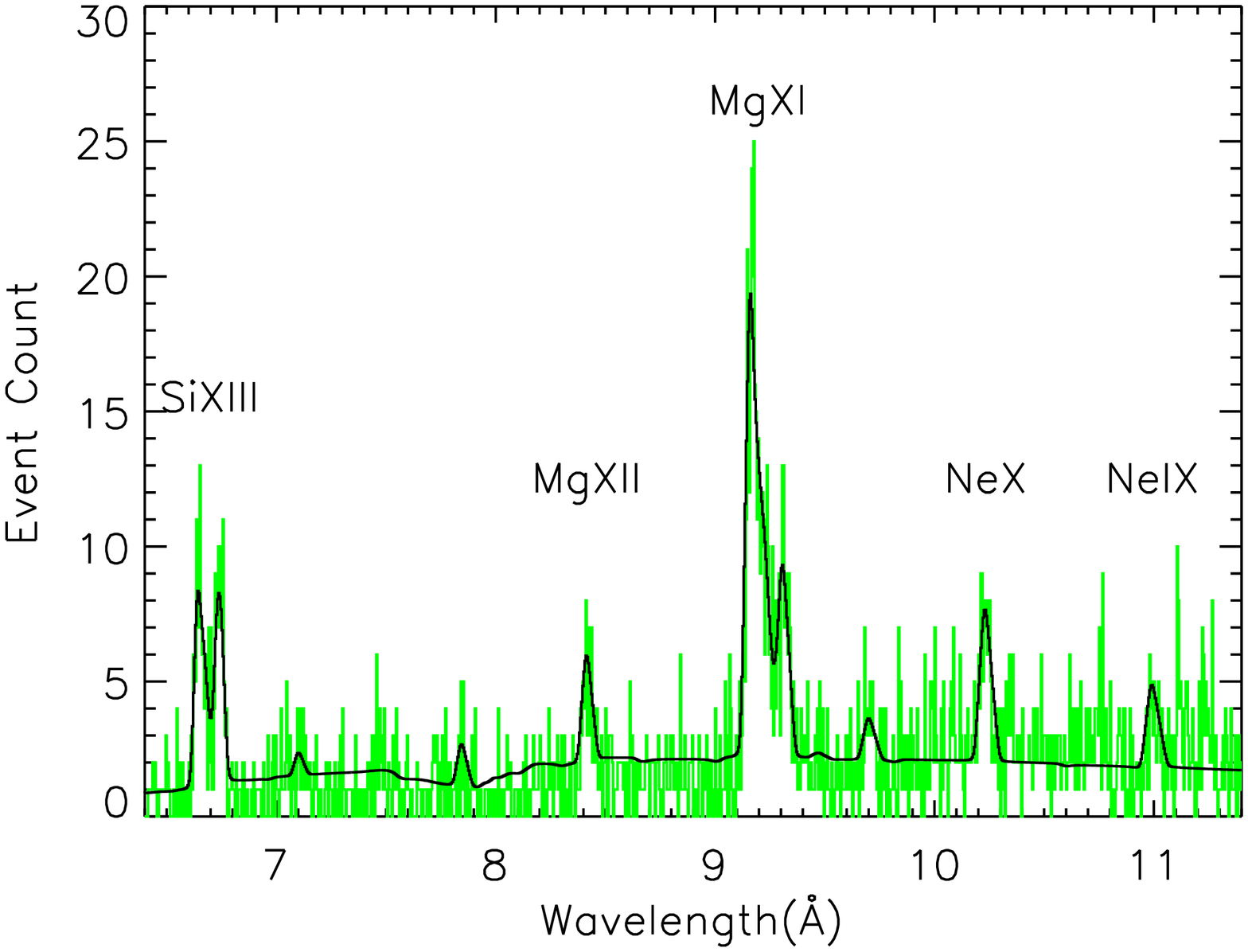}}} \\
\end{tabular}
\end{center}
\caption{\label{Figure:RGS:SingleTriLine:fits}
    Selected comparisons of high-resolution X-ray data with the best-fit single mean {\tt TriLine} profile 
    model used for all the lines with the 1st order RGS1 in red, RGS2 in blue and the combined MEG 1st order 
    in green. The common model is shown by the smooth black line. The \ion{O}{viii} Ly$\alpha$~line
    was affected by bad pixels in RGS1.}
\end{figure}

The RGS and HETG spectra, which were taken over two years apart, agree
well with each other both in the shapes of the lines and in overall
luminosity to within $10$ or $20\%$
as shown in Table~\ref{Table:TriLine:normalisations}.
The single line-profile fits showed very similar red and blue velocities
with a significant asymmetry due to a blue-shift of the line peak. 
The velocity errors are small by virtue of combining
several hundred independent estimates of a single profile
from the lines of different spectra.
The statistical errors take no account of any systematic differences
between lines
although the model-independent and statistical
analyses reported below and the comparisons between data and model
shown in Fig.~\ref{Figure:RGS:SingleTriLine:fits} all suggest
that the common profile gives a good representation.
It establishes that the peak of the X-ray lines in \hbox{$\zeta$~Orionis}
is significantly blue-shifted by about $-300{\rm km}~{\rm s}^{-1}$,
similar though not as much as the shift
in \hbox{$\zeta$~Puppis}.
The values of blueV and redV are
$80\%$ of \hbox{$\zeta$~Orionis}'s terminal
velocity of $v_{\infty}=2100{\rm km}~{\rm s}^{-1}$
or 90\% of the preferred value of \cite{PBH:1990}.
They are accurate enough to compare with any of the
velocity parameters used
to describe the complex shape of UV P Cygni profiles
that trace cool pre-shock material.

In order to help comparison with earlier work and allow an
assessment of the significance of the line asymmetry,
we also implemented a {\tt SkewLine} model of
a Gaussian line with different blue and red velocity half widths.
Table~\ref{Table:SkewLine:model} shows the results of a number of
{\tt TriLine} and {\tt SkewLine} fits subject to different
shift and symmetry constraints. The fits are of the same data
and can thus be compared quantitatively with the likelihood
C-statistic. The best of the eight fits in the Table is the blue-shifted
{\tt SkewLine} model which gives very similar parameters
for the mean velocity distribution of line-emitting material
as the best {\tt TriLine} model. It is clear, in particular
from comparison with the best-fit symmetrical unshifted models,
that the mean line is very significantly skewed and blue-shifted.
The symmetrical Gaussian results agree well with those reported earlier
by \cite{M:2002} for the {\it Chandra} data.

Although the asymmetrical {\tt SkewLine} gives a better fit,
we prefer the more convenient {\tt TriLine} parameterisation that is
easier to relate to other observable quantities, velocity
measurements and the line models calculated in the literature.
The superiority of the {\tt SkewLine} model may be related to
ability of its wings to help account for weaker neighbouring lines
not included in the model line list.
In any case, longer-exposure data would allow a more reliable statistical comparison.
With the single shifted, asymmetric {\tt TriLine} model of Table~\ref{Table:TriLine:model}
for \hbox{$\zeta$~Orionis}, the line fluxes listed in
Table~\ref{Table:Line:fluxes} were determined,
where no corrections for absorption were made.
Here too, the measurements in common agree well with Miller's.

\begin{table}[ht]
\caption{\label{Table:TriLine:model}
         \hbox{$\zeta$~Orionis} best-fit emission-line velocity parameters and underlying continuum
         for the simultaneous multiple common-profile {\tt TriLine} {\tt XSPEC}
         model fit to the {\it XMM-Newton} RGS 1st and
         2nd order spectra of 2002-09-15 and
         {\it Chandra} MEG and HEG $\pm1$ order spectra of 2000-04-08.}
\begin{center}
\begin{tabular}{lr@{$\pm$}ll}
\hline
\multicolumn{4}{c}{{\tt TriLine} velocities}                                               \\
\hline
blueV                & $-1642$ & $22$                         & ${\rm km}~{\rm s}^{-1}$    \\
centralV             &  $-302$ & $29$                         & ${\rm km}~{\rm s}^{-1}$    \\
redV                 & $+1646$ & $26$                         & ${\rm km}~{\rm s}^{-1}$    \\
\hline
\multicolumn{4}{c}{bremsstrahlung continuum}                                               \\
\hline
${\rm N}_{\rm H}$    & \multicolumn{2}{c}{$2.5\times10^{20}$} & ${\rm cm}^{-2}$            \\
kT                   & $0.494$ & $0.007$                      & keV                        \\
normalisation        & $5.66$  & $0.14\times10^{-3}$          &                            \\
\hline
\multicolumn{4}{c}{C-statistic=22862.4 using 26281 PHA bins} \\
\hline
\end{tabular}
\end{center}
\end{table}

\begin{table}[ht]
\caption{\label{Table:TriLine:normalisations}
         Best-fit normalisation parameters for the model of Table~\ref{Table:TriLine:model}.
         {\tt TriLine} model fit to
         high-resolution spectra of \hbox{$\zeta$~Orionis} taken
         with {\it XMM-Newton} on 2002-09-15 and {\it Chandra} on 2000-04-08.}
\begin{center}
\begin{tabular}{lrrr@{$.$}l@{$\pm$}l}
\hline
Instrument & Order & Counts & \multicolumn{3}{c}{Relative Normalisation} \\
\hline
RGS1 & $-1$ &  15136 & $1$&$02$ & $0.01$      \\
RGS2 & $-1$ &  14133 & \multicolumn{3}{l}{1}  \\
RGS1 & $-2$ &   2771 & $1$&$13$ & $0.01$      \\
RGS2 & $-2$ &   2507 & $1$&$00$ & $0.01$      \\
MEG  & $-1$ &   5556 & $1$&$11$ & $0.02$      \\
MEG  & $+1$ &   3568 & $1$&$03$ & $0.02$      \\
HEG  & $-1$ &   1129 & $1$&$18$ & $0.03$      \\
HEG  & $+1$ &    839 & $1$&$12$ & $0.03$      \\
\hline
\end{tabular}
\end{center}
\end{table}

\begin{table*}[ht]
\caption{\label{Table:SkewLine:model}
         Best-fit emission-line velocity parameters in ${\rm km}~{\rm s}^{-1}$
         for variously constrained {\tt TriLine} and blue and red half-Gaussian {\tt SkewLine} model fits to the grating spectra
         of \hbox{$\zeta$~Orionis} from the {\it XMM-Newton} RGS and the {\it Chandra} MEG and HEG.}
\begin{center}
\begin{tabular}{rr@{$\pm$}lr@{$\pm$}lr@{$\pm$}lrl}
\hline
                     & \multicolumn{2}{c}{blueV}
                                      & \multicolumn{2}{c}{centralV}
                                                       & \multicolumn{2}{c}{redV} 
                                                                        & \multicolumn{1}{c}{C-stat}
                                                                                  & constraint \\
\hline
{\tt TriLine}        & $-1642$ & $22$ &  $-302$ & $29$ & $+1646$ & $26$ & 22864.4 & Best fit of Table~\ref{Table:TriLine:model} \\
{\tt TriLine}        & $-1641$ & $15$ &  $-303$ & $26$ & $+1641$ & $15$ & 22865.9 & $|$redV$|$=$|$blueV$|$ \\
{\tt TriLine}        & $-1774$ & $21$ &  \multicolumn{2}{c}{0}
                                                       & $+1492$ & $21$ & 22973.1 & centralV=0 \\
{\tt TriLine}        & $-1684$ & $15$ &  \multicolumn{2}{c}{0}
                                                       & $+1684$ & $15$ & 23053.2 & centralV=0,$|$redV$|$=$|$blueV$|$ \\
\hline
\hline
                     & \multicolumn{2}{c}{blueW}
                                      & \multicolumn{2}{c}{centralV}
                                                       & \multicolumn{2}{c}{redW} 
                                                                        & \multicolumn{1}{c}{C-stat}
                                                                                  & constraint \\
\hline
{\tt SkewLine}       &   $650$ & $25$ &  $-318$ & $29$ &   $991$ & $27$ & 22816.7 & \\
{\tt SkewLine}       &   $811$ & $ 9$ &  $-126$ & $10$ &   $811$ & $ 9$ & 22879.4 & redW=blueW \\
{\tt SkewLine}       &   $880$ & $12$ &  \multicolumn{2}{c}{0}
                                                       &   $738$ & $12$ & 22947.6 & centralV=0 \\
{\tt SkewLine}       & $  841$ & $10$ &  \multicolumn{2}{c}{0}
                                                       & $  841$ & $10$ & 23038.4 & centralV=0, redW=blueW \\
\hline
\end{tabular}
\end{center}
\end{table*}

\clearpage

\subsection{The composite model-independent line profile \label{SubSection:CompositeLine}}

\begin{figure}
\begin{center}
\begin{tabular}{c}
\rotatebox{90}{\scalebox{0.25}{\includegraphics{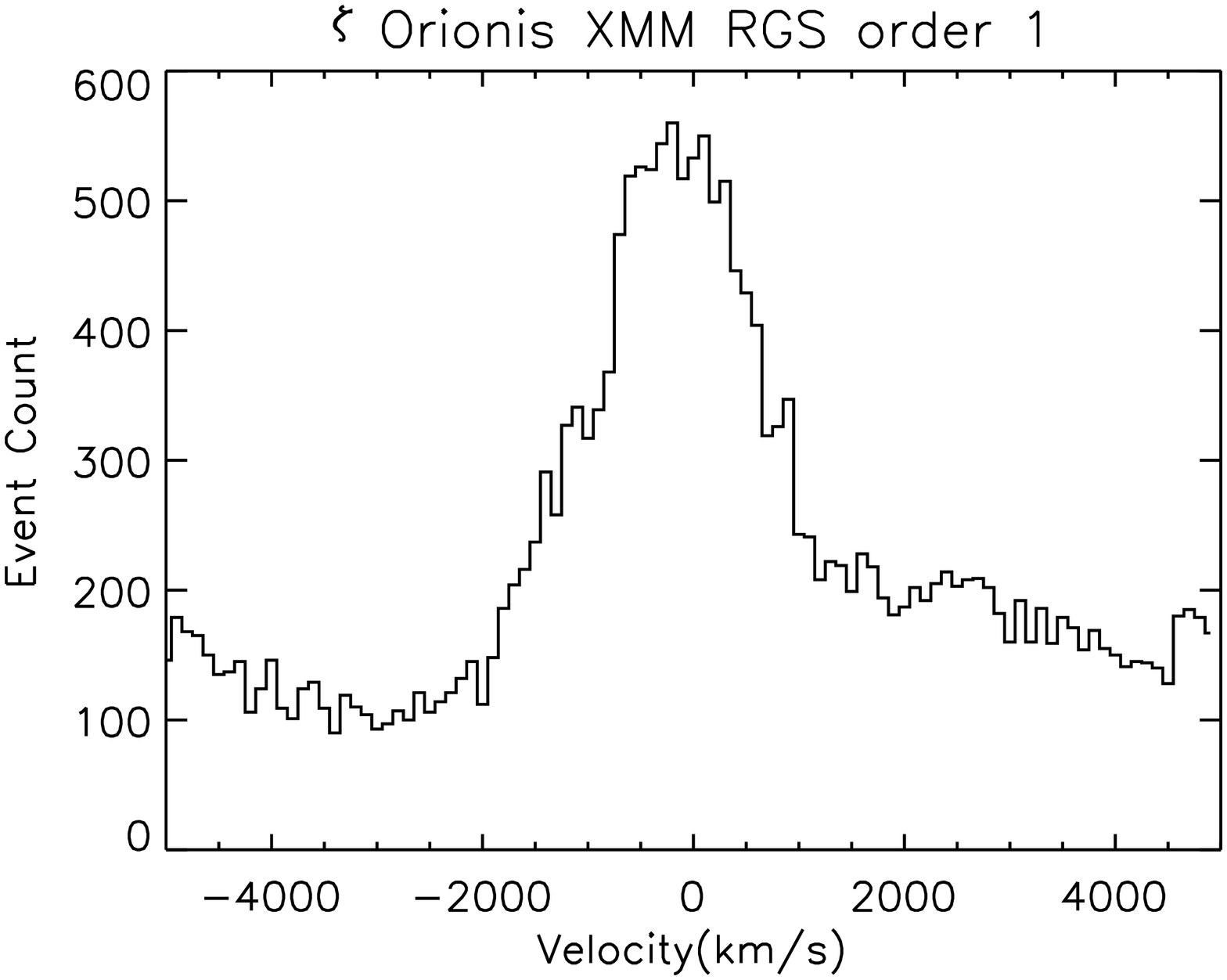}}} \\
\rotatebox{90}{\scalebox{0.25}{\includegraphics{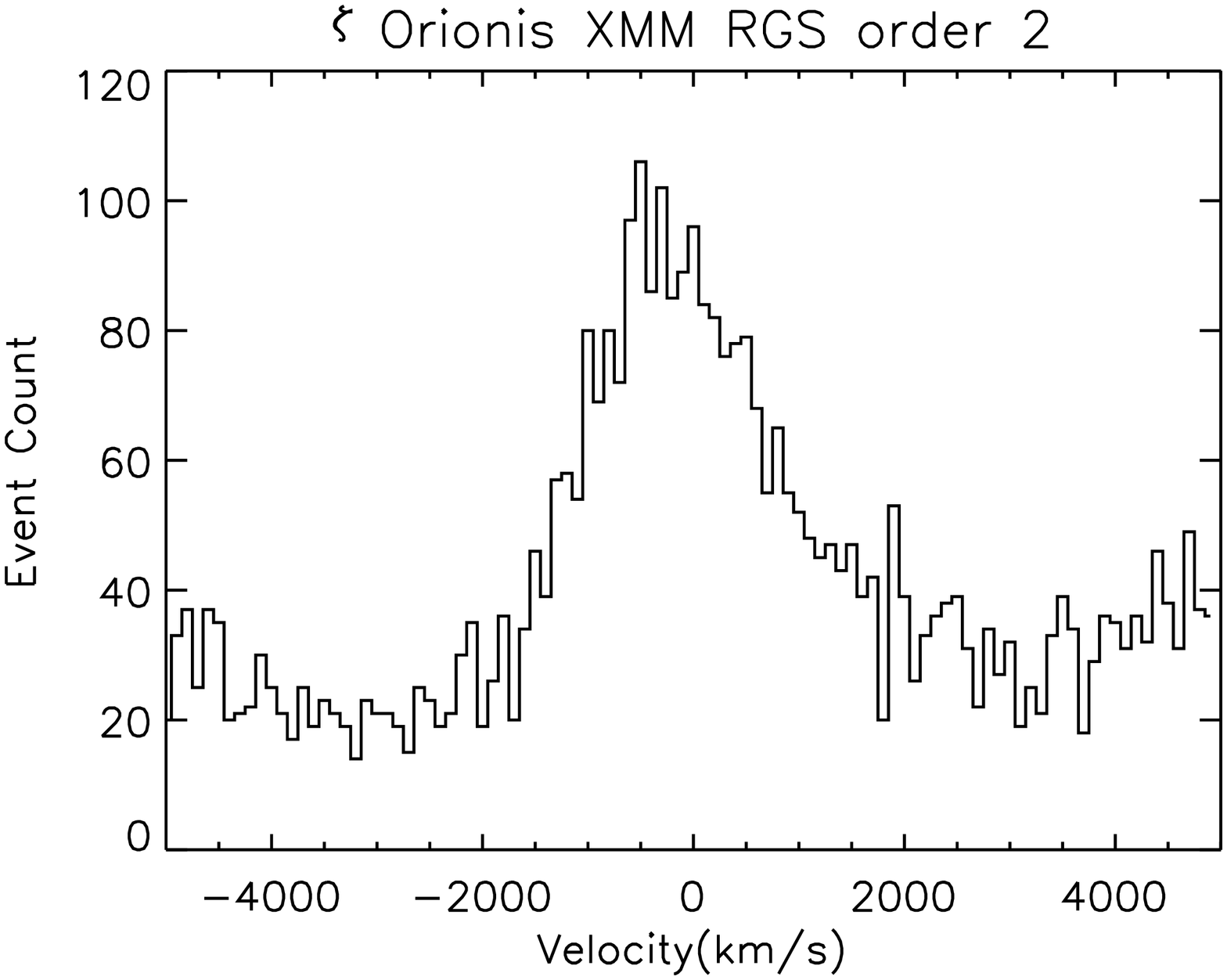}}} \\
\rotatebox{90}{\scalebox{0.25}{\includegraphics{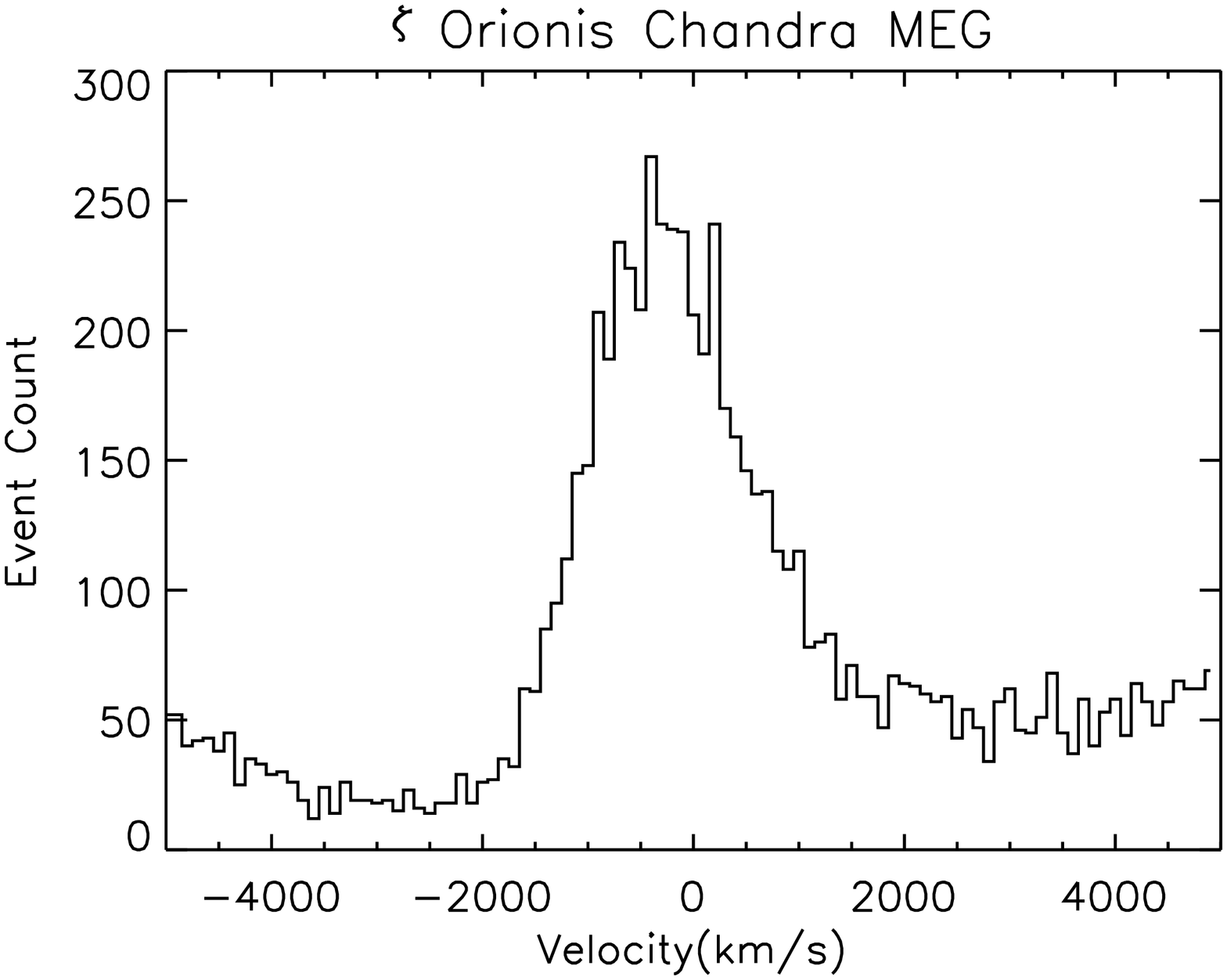}}}    \\
\rotatebox{90}{\scalebox{0.25}{\includegraphics{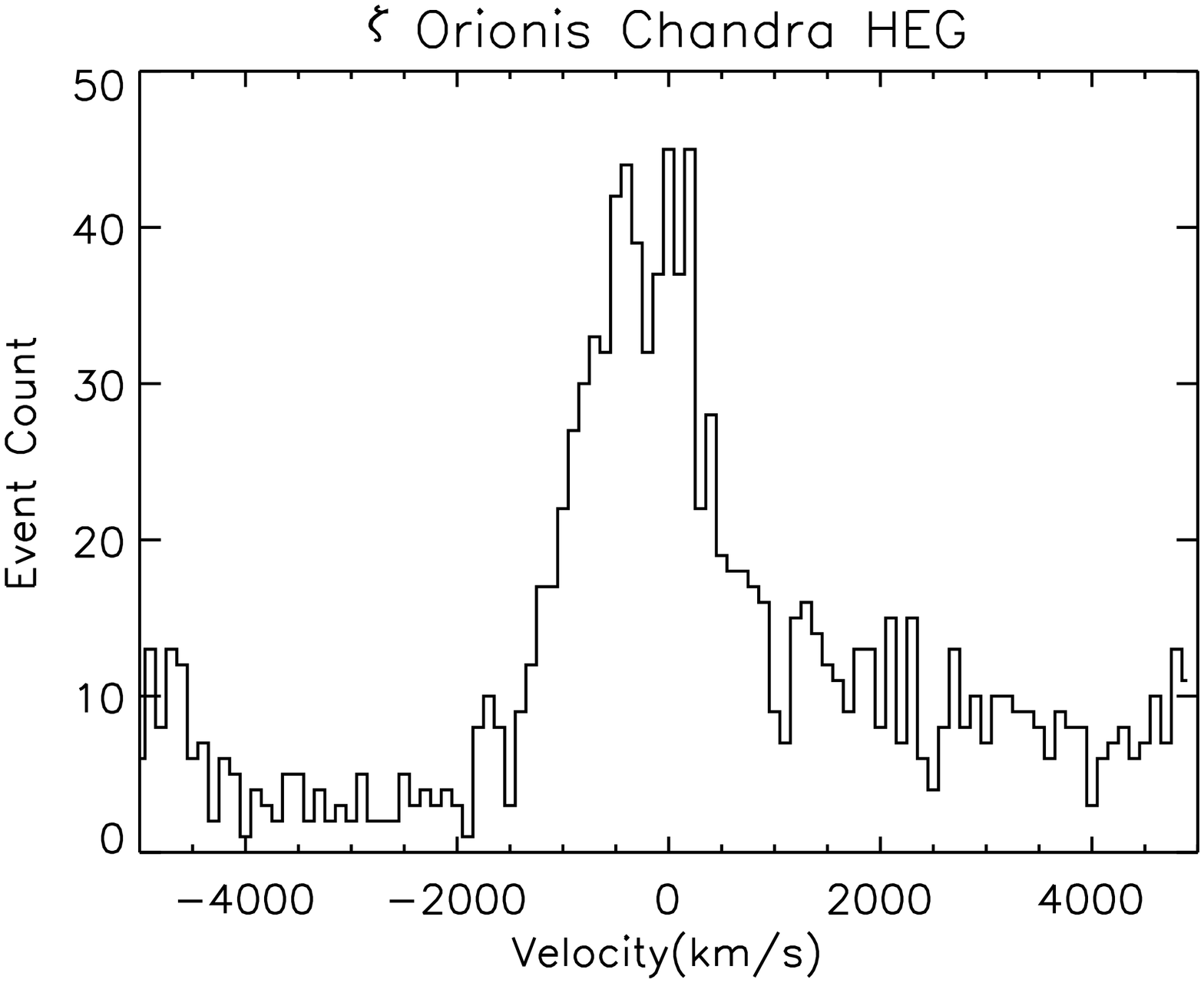}}}    \\
\end{tabular}
\end{center}
\caption{\label{Figure:RGS:Chandra:Line}
    Composite X-ray line profiles of \hbox{$\zeta$~Orionis} calculated from {\it XMM-Newton} RGS and {\it Chandra} HETG grating data.
    The instrumental resolution improves
    down the page so that the observed profile becomes a better approximation to the actual velocity dispersion of
    X-ray emitting material. The profile at $v>+1000~{\rm km}~{\rm s}^{-1}$ is distorted by the intercombination lines of the He-like
    triplets. \hbox{$\zeta$~Orionis}'s terminal velocity is $v_{\infty}=2100{\rm km}~{\rm s}^{-1}$.}
\end{figure}

The important model-independent question arises of whether any
X-ray gas was observed at velocities greater than $v_{\infty}$.
We have calculated the distribution of all the detected events in velocity
space by stacking all the individual line velocity distributions.
Any continuum or background photons should form a smooth background
in the composite velocity space. The resultant profiles from both
{\it XMM} and {\it Chandra} data are shown in
Fig.~\ref{Figure:RGS:Chandra:Line}. It should be emphasized that
this plot has limitations in that it includes the
instrumental resolutions and neglects sensitivity variations, such
as bad pixels, that are properly taken into account in the response
matrices; the RGS 1st order profile, for example, was affected by
CCD columns missing near the centre of the \ion{O}{viii}
Ly$\alpha$~$\lambda18.967$. As these defects have nothing to do
with the distribution of moving gas in
\hbox{$\zeta$~Orionis}, we should be able assess reasonably
objectively from the ensemble of lines the range of velocities
present. Taking into account the differing resolving power in the
four spectra, they agree well with each other and with the {\tt
TriLine} velocities reported in
Table~\ref{Table:TriLine:model}. The blue wing is sharp
and joins the weak underlying continuum before the
terminal velocity is reached. The red wing, on the other hand,
flattens above about $+1000~{\rm km}~{\rm s}^{-1}$ because of the
strength of the He-like triplet intercombination lines. It is quite
safe to conclude from Fig.~\ref{Figure:RGS:Chandra:Line} that there
was no X-ray emitting gas moving at or above the wind terminal
velocity.

\subsection{An assessment of line-profile variations}

The sharpness of the composite line profile is consistent with the idea
suggested by Fig.~\ref{Figure:RGS:CVI:MEG:NeX}
that all ions have much
the same shape. In order to check this more rigorously, we have
estimated individual best-fit line profiles for
the most prominent ions as reported in Table~\ref{Table:TriLine:ions}. As well as
showing the individual ion {\tt TriLine} parameters, the statistical significance of the
differences may be judged using $\Delta$C, the decrease in the value of the C-statistic
gained by freeing an ion's {\tt TriLine} parameters. Under the null hypothesis of
a common line profile, $\Delta$C should be distributed as $\chi_3^2$ for the 3 extra
degrees of freedom. The variations between ions are small,
amounting to less than about \hbox{$200~{\rm km}~{\rm s}^{-1}$}
from one ion to another, but probably real.
Nevertheless, the assumption of a single profile looks like a sensible working hypothesis.

\begin{figure}
\begin{center}
\rotatebox{90}{\scalebox{0.3}{\includegraphics{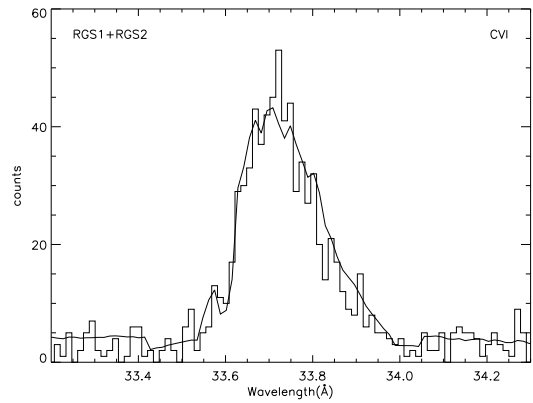}}} \\
\end{center}
\caption{\label{Figure:CVI:data:model}
    The \ion{C}{vi} Ly$\alpha$ line of \hbox{$\zeta$~Orionis} observed with the {\it XMM-Newton} RGS
    shown by the histogram compared with the mean {\tt TriLine} model profile of
    the joint fit of Table~\ref{Table:TriLine:model} to all the lines in the spectrum.}
\end{figure}

It has been argued by both \cite{KLC:etal:2001} and \cite{CMWMC:2001}
that the \ion{N}{vii} Ly$\alpha$~line in \hbox{$\zeta$~Puppis}
has a much wider
flat-topped profile, significantly different from the other lines, so that a single shape does not
apply. However, the RGS data of \hbox{$\zeta$~Puppis} suggest 
that the \ion{N}{vii} line might not be unusually wide
if account is taken of blending with \ion{N}{vi} He$\beta$:
the \ion{N}{vi} He-like triplet is strong and in common with \ion{N}{vii} Ly$\beta$
and \ion{C}{vi} Ly$\alpha$ is not obviously unusually broad.
We intend to discuss \hbox{$\zeta$~Puppis} in detail in a future article.
In \hbox{$\zeta$~Orionis}, the evidence from Table~\ref{Table:TriLine:ions}
and Fig.\ref{Figure:CVI:data:model} is that
the longest-wavelength \ion{C}{vi} line is, if anything, narrower
than average rather than broader.

\begin{table}[ht]
\caption{\label{Table:TriLine:ions}
         Best-fit {\tt TriLine} model velocity parameters for individual prominent 
         ions in the X-ray spectrum of \hbox{$\zeta$~Orionis} compared
         with the single common line model. The quoted 1$\sigma$ errors
         were estimated using the {\tt XSPEC> error 1.} command }
\begin{center}
\begin{tabular}{lr@{$\pm$}lr@{$\pm$}lr@{$\pm$}lrr}
\hline
\multicolumn{1}{c}{ion}
                     & \multicolumn{2}{c}{blueV}
                                       & \multicolumn{2}{c}{centralV}
                                                         & \multicolumn{2}{c}{redV}
                                                                            & C-stat  & $\Delta$C \\
                     & \multicolumn{2}{c}{(${\rm km}~{\rm s}^{-1}$)}
                                       & \multicolumn{2}{c}{(${\rm km}~{\rm s}^{-1}$)}
                                                         & \multicolumn{2}{c}{(${\rm km}~{\rm s}^{-1}$)}
                                                                            &         &      \\
\hline
All ions             & $-1642$ &  $22$ & $-302$ &  $29$  & $+1646$ &  $26$  & 22862.4 &      \\
\hline
\ion{C}{vi}           & $-1719$ & $135$ & $-270$ & $137$  & $+1403$ & $130$  & 22855.6 &  6.8 \\
\ion{N}{vii}          & $-1705$ & $ 55$  & $-111$  & $240$  & $+1065$ & $324$  & 22859.8 &  2.6 \\
\ion{O}{vii}          & $-1349$ & $ 72$  & $-509$ & $ 83$   & $+1835$ & $ 76$   & 22844.9 & 17.5 \\
\ion{O}{viii}         & $-1656$ & $ 43$  & $-258$ & $ 59$   & $+1482$ & $ 46$   & 22838.7 & 23.7 \\
\ion{Fe}{xvii}       & $-1647$ & $ 42$  & $-421$ & $ 62$   & $+1845$ & $ 59$   & 22848.9 & 13.5 \\
\ion{Ne}{ix}         & $-1653$ & $ 77$  & $-211$  & $ 79$   & $+1508$ & $102$  & 22859.9 &  2.5 \\
\ion{Ne}{x}          & $-1730$ & $106$ & $-327$ & $126$  & $+1735$ & $103$  & 22861.0 &  1.4 \\
\ion{Mg}{xi}         & $-1536$ & $153$ & $ +33$ & $121$  & $+1487$ & $144$  & 22850.1 & 12.3 \\
\ion{Mg}{xii}        & $ -721$  & $365$  & $-662$ & $227$ & $+1514$ & $471$  & 22855.1 &  7.3 \\
\ion{Si}{xiii}         & $-1449$ & $124$  & $ +48$ & $240$  & $+1421$ & $188$  & 22856.4 &  6.0 \\
\hline
\end{tabular}
\end{center}
\end{table}

\clearpage

\section{The X-ray continuum in \hbox{$\zeta$~Orionis}\label{Section:Continuum}}

\begin{figure*}
\begin{center}
\rotatebox{90}{\scalebox{0.5}{\includegraphics{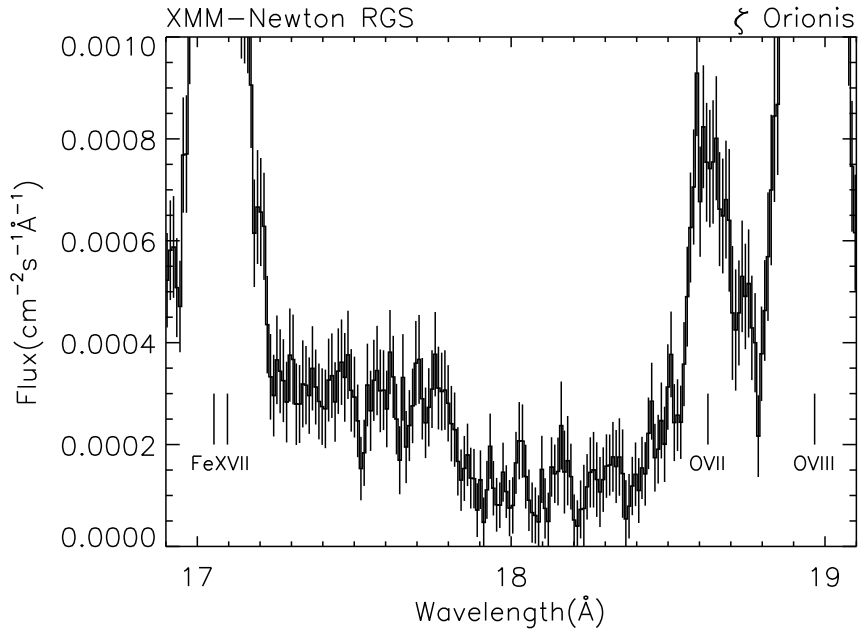}}}
\end{center}
\caption{\label{Figure:Spectrum:17-19}
    The {\it XMM-Newton} RGS spectrum of \hbox{$\zeta$~Orionis} between the lines of
    \hbox{\ion{Fe}{xvii}~$\lambda\lambda17.053,17.098$}
    and \hbox{\ion{O}{vii}~$\lambda18.627$} and \hbox{\ion{O}{viii}~$\lambda\lambda18.967,18.972$}.
    Some of the weak lines between about 17.8 and 18.4 \AA~are probably
    transitions to the excited \hbox{$2s2p^6$} level of
    \ion{Fe}{xviii}, which appear in the ATOMDB and CHIANTI line lists with uncertain
    wavelengths. The level of the continuum looks to be less
    than about \hbox{$5\times10^{-5}{\rm cm}^{-2}{\rm s}^{-1}{\rm \AA}^{-1}$} in this part of the spectrum.}
\end{figure*}

Although a semi-empirical continuum component was added to the line models described above,
the real nature and strength of the continuum is an important matter
as it is a direct diagnostic of the density and temperature of plasma electrons.
In common with other O-stars, the continuum is weak.
Not only are lines responsible for the most obvious features in the
spectrum in Fig.~\ref{Figure:RGS:spectrum}, but increasing numbers of weaker lines seem to make up
most of the emission observed.
Fig.~\ref{Figure:Spectrum:17-19} shows details of the spectrum 
between 17 and 19~\AA.
The low flux between about 17.8 and 18.5~\AA,
that contrasts to the plateau immediately
bluewards, is common to the handful of O stars observed so far
but still shows features that could plausibly be identified with
upper level transitions of \ion{Fe}{xviii} for which
no observed wavelengths are available in the line lists.
The flux reached similarly low but again not especially smooth values between 20 and 21~\AA,
where plausible identifications are with lines of \ion{N}{vii} and \ion{Ca}{xvii}.

The high-resolution X-ray spectra of well known relatively high-temperature active cool stars, such as AB~Dor and HR~1099,
show obvious strong continuum emission, although this is not the case 
with cooler stars such as Procyon \citep{R:etal:2002}, where the continuum is notably weak.
We considered global empirical line+continuum models for \hbox{$\zeta$~Orionis}
with more than the 85 lines already included but these soon
became impractical as the number of weak lines increases exponentially and the wavelengths are uncertain.
If a continuum is present in \hbox{$\zeta$~Orionis}, its flux near 18\AA\ is
less than about \hbox{$5\times10^{-5}{\rm cm}^{-2}{\rm s}^{-1}{\rm \AA}^{-1}$}.
A long exposure would help to constrain it with more confidence.

\clearpage

\section{Origin of the X-rays from \hbox{$\zeta$~Orionis}}

\begin{figure*}
\begin{center}
\begin{tabular}{c}
\rotatebox{90}{\scalebox{0.50}{\includegraphics{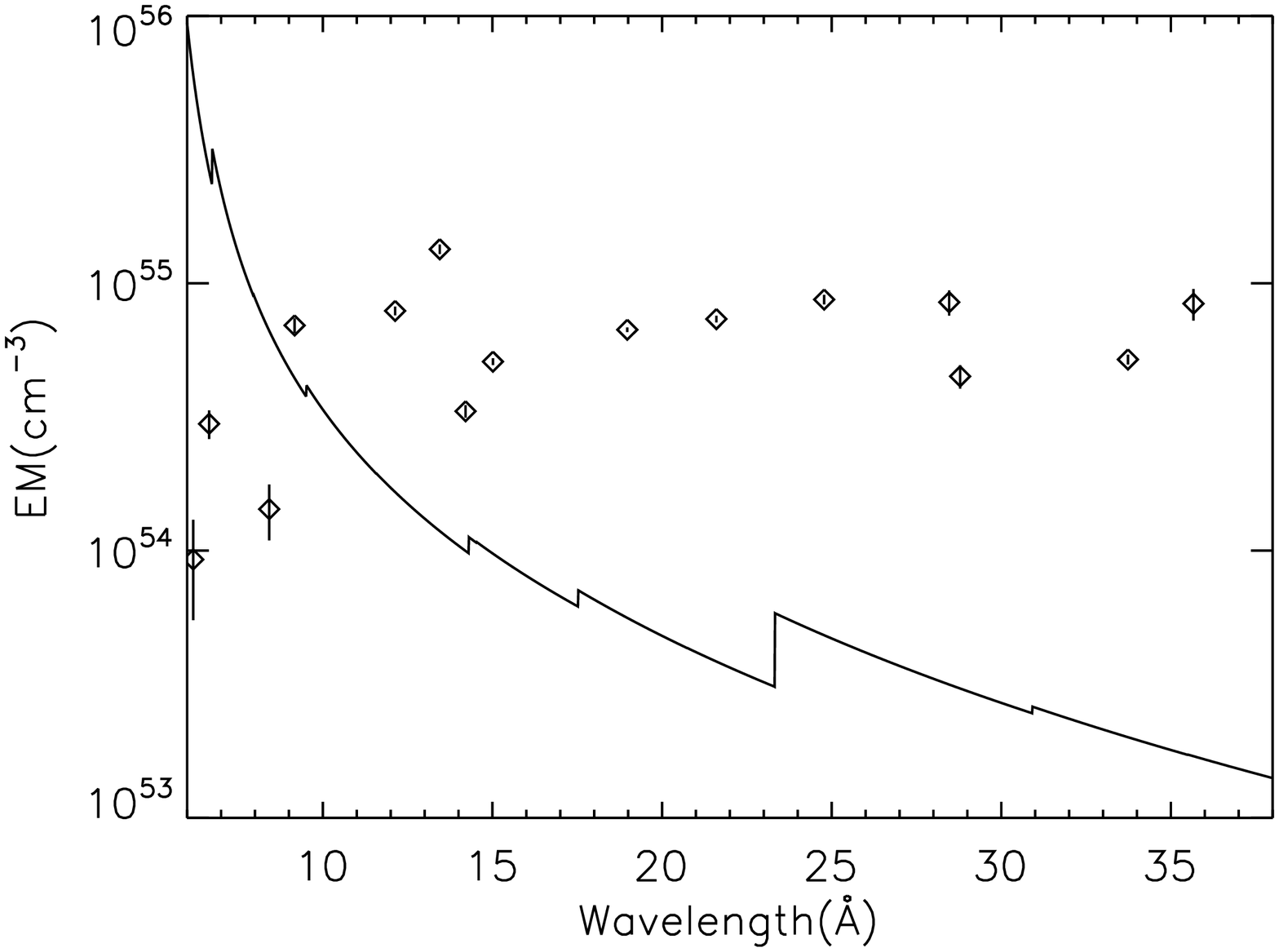}}} \\
\rotatebox{90}{\scalebox{0.50}{\includegraphics{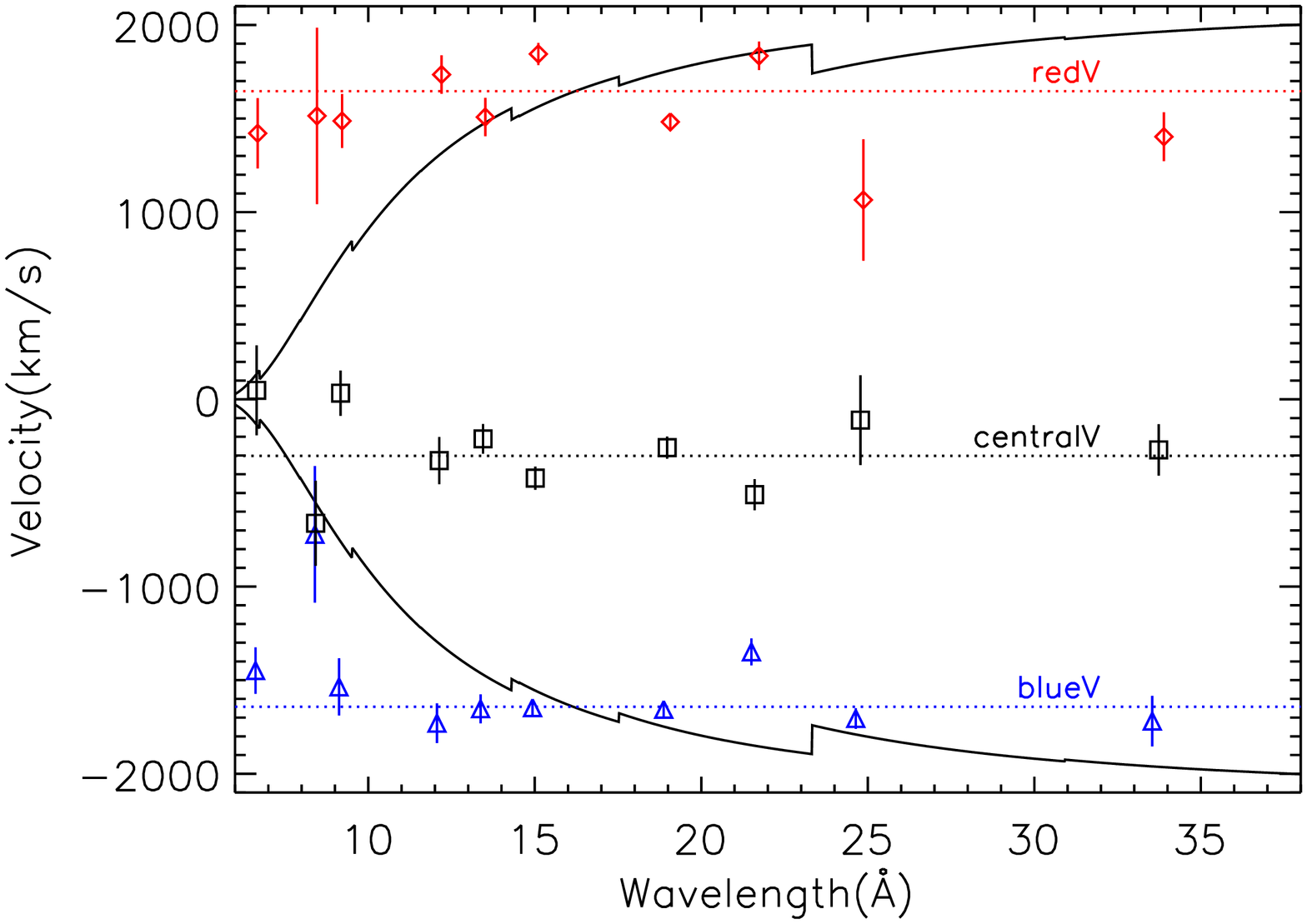}}} \\
\end{tabular}
\end{center}
\caption{\label{Figure:LineProperties}
    Observed properties of the X-ray lines of \hbox{$\zeta$~Orionis}. In the top panel are plotted the line
    emission measures, $EM=4{\pi}d^2(f/\epsilon_{\rm max})$, calculated from the measured fluxes, $f$,
    of Table~\ref{Table:Line:fluxes},
    the maximum solar-abundance APED emissivities, $\epsilon_{\rm max}$, and the distance, $d$, to the star.
    The small error bars reflect the high precision with which the fluxes are known.
    Below are plotted the measured {\tt redV}, {\tt centralV} and {\tt blueV} parameters of the {\tt TriLine} models
    of individual line profiles from Table~\ref{Table:TriLine:ions},
    with the corresponding mean values shown by the horizontal lines. Also shown as solid
    lines are two illustrative quantities calculated for $R_{\tau=1}(\lambda)$, the radius of the $\tau=1$
    X-ray absorption optical depth for a $\beta=0.8$ velocity law. At the top is $EM(>R_{\tau=1}(\lambda))$,
    the emission measure
    outside $R_{\tau=1}(\lambda)$, scaled down to overlap the data. Below is ${\pm}v(R_{\tau=1}(\lambda))$,
    the range of velocities offered by the accelerating wind at that radius.}
\end{figure*}

The measured fluxes and velocity widths of the lines in the X-ray spectrum of \hbox{$\zeta$~Orionis} are plotted
in Fig.~\ref{Figure:LineProperties}. The fluxes have been converted to emission measures using the
ATOMDB\footnote{http://cxc.harvard.edu/atomdb/}
maximum emissivities, similar to the plots against temperature for \hbox{$\zeta$~Puppis} by
\citet{KLC:etal:2001} and \hbox{$\delta$~Orionis} by \citet{MCWMC:2002}.
In common with those and other stars, the X-ray emission measures are three or four orders
of magnitude below that available in the wind as a whole
and show that only a small fraction of the wind material is involved.

The remarkable thing about
Fig.~\ref{Figure:LineProperties} is the lack of any substantial variation with wavelength of either emission
measure or velocity width, in contrast to the rough expectations of the instability-driven shock
model also plotted there.
These have been calculated
for the parameters of \hbox{$\zeta$~Orionis} in Table~\ref{Table:zO:data}
assuming a velocity law \hbox{$v(r)=v_{\infty}(1-R_*/r)^{\beta}$} with $\beta=0.8$.
In the development of standard model
following \cite{WC:2001}'s first discussion of
\hbox{$\zeta$~Orionis}'s {\it Chandra} grating spectrum,
X-rays at shorter wavelengths are thought to emerge from deeper into the slower, denser material
of the acceleration zone of the wind as the opacity decreases,
with the corresponding expected increase in the emission measure plotted in 
Yet the measurements show little or no evidence of this. Rather they give quantitative expression to the
impression of Fig~\ref{Figure:RGS:spectrum} that the spectrum is more-or-less optically thin.
If this is the case, then the suggestion is that X-rays are not from the dense central regions of the acceleration
zone but from further out in the terminal velocity region, although a more precise location is not known.
How this might be reconciled with the apparently firm conclusions concerning the location
of the X-ray emission deduced from analysis of the He-like triplets is one of the topics
discussed below.

In the following sections, an attempt is made to discuss in some detail the feasibility and consequences of the
possibility that X-rays arise in the wind's terminal velocity regime.
It should be stressed that, in common with any ideas, success or failure
should be judged by three things:
consistency with the law of physics;
ability to reproduce current data, such as those in Fig.~\ref{Figure:LineProperties}; and
ability to allow predictions to be made that can be tested against future observations.

To begin with, it is tempting to interpret the simultaneous presence in the
spectrum of lines of a range of ionization potentials as proof of
the presence of a range of temperatures, lending support to the
proposal that cooling rather than isothermal shocks are responsible
for the X-ray emission. For \hbox{$\zeta$~Orionis}, using the
well-known methods offered, for example, by
{\tt XSPEC} 
or
{\tt SPEX}\footnote{http://www.sron.nl/divisions/hea/spex/},
it is possible to make a good
synthesis of the spectrum with a multi-temperature model involving
gas in collisional ionization equilibrium between about 1MK and 7MK
involving either a limited number of discrete temperature components
or a continuous DEM distribution, like that suggested by Fig.~\ref{Figure:LineProperties}.
However, before accepting at face
value the temperatures that this would imply, it is worth
considering the general physical principles discussed by
\cite{ZR:2002} that govern shocks and how these might affect their
development in the particular circumstances of a stellar wind,
especially if the shocks occur in the relatively low density
terminal velocity region.
Recently \cite{PCSW:2005} did this type of analysis for the
binary-system colliding-wind shocks in WR~140. Similar considerations
for single hot stars lead to some interesting conclusions.

\clearpage

\subsection{The nature of shocks in O-star winds}

\begin{figure}
\begin{center}
\rotatebox{90}{\scalebox{0.3}{\includegraphics{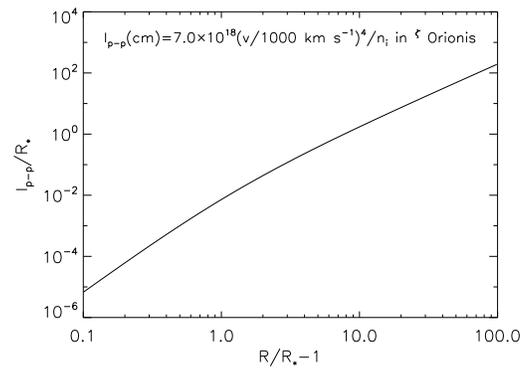}}}
\end{center}
\caption{\label{Figure:ion-ion:mean-free-path}
    The mean free path for ion-ion Coulomb collisions in the wind of \hbox{$\zeta$~Orionis}.}
\end{figure}

An O-star wind is a plasma flow in which particle interactions are
long-range Coulomb collisions. These have a key role in the wind
acceleration mechanism in normal circumstances as Coulomb coupling
is responsible for redistribution of momentum from the small
minority of UV-driven ions to the rest of the flow
\citep[e.g.][]{KP:2000}. In more extreme conditions, \cite{ZR:2002}
described the particular properties of plasma shock waves that
result from the slow character of the Coulomb energy exchange
among ions and even slower between ions and electrons.
Because most of the kinetic energy of the flow is in the ions,
the most important quantity that
determines shock development is the ion-ion collisional mean-free path, which
can be estimated as the distance required to deflect by $90^o$
through successive Coulomb collisions a proton or He nucleus moving
with the velocity, \hbox{$v=v_8\times1000{\rm km}~{\rm s}^{-1}$}.
Following \cite{DM:1993} after \cite{S:1962}, the mean-free-path is
\begin{equation}\label{Equation:lii}
l_{\rm i-i} \sim 7.0\times10^{18}{v_8}^4/n_{\rm i}~{\rm cm}
\end{equation}
where $\hbox{$n_{\rm i}$}$ is the ion density.
Fig.~\ref{Figure:ion-ion:mean-free-path} shows how the ion-ion
collisional mean-free path varies throughout the wind of
\hbox{$\zeta$~Orionis}.

Close to the photosphere, where the density is high and the velocity
low, $\hbox{$l_{\rm i-i}$}$ is small, but then increases rapidly
with the increasing velocity and falling density. At about $3R_*$
above the stellar surface in the acceleration zone,
$\hbox{$l_{\rm i-i}$}\approx0.1R_*$ while at $10R_*$, $\hbox{$l_{\rm
i-i}$}\approx R_*$. The implications are profound. If strong shock
discontinuities are to develop at all in an O-star wind then some
other dissipation mechanism than collisions must come into operation:
they must be collisionless shocks. As
discussed by \cite{DM:1993}, for example, collisionless shocks
probably account for the majority of shocks observed in cosmic
plasmas, which are of predominantly low density and thus of long
collisional mean-free-path. Familiar examples are the interaction
between the solar wind and the earth's magnetosphere and the blast
wave of a SNR sweeping through the interstellar medium; more
relevant might be the colliding-wind shocks of WR~140
\citep{PCSW:2005}. In all these cases, shock dissipation occurs
though collective plasma processes involving a magnetic field in
which a small characteristic dissipation length is set by the ion Larmor
radius. The development of shocks in single stars and the production
of X-rays is thus likely to be a phenomenon controlled by magnetic fields
at large within the wind.

\clearpage

\subsection{Post-shock relaxation and the electron temperature}

As described by \cite{ZR:2002}, a property of plasma shock
transitions is that ions are heated to high temperatures while
electrons initially remain relatively cold. This is certain for shocks in media dense
enough for Coulomb collisional dissipation to operate but is
probably also largely true in collisionless shocks, according to the recent
review by \cite{R:2006}.
Although the amount of electron shock heating by plasma turbulence
above the minimum provided by
adiabatic heating remains a matter of speculation,
Rakowski showed that the evidence from SNR shocks is that electron
heating decreases with shock velocity
and should be small for the few thousand ${\rm km}~{\rm s}^{-1}$
typical of stellar-wind terminal velocities.
It is in the subsequent flow through the post-shock relaxation layer
that energy is transferred
from ions to electrons as the plasma moves towards equilibrium.
Naturally enough,
for the use of equilibrium plasma models to be justified in the post-shock gas,
equilibrium between ions and electrons needs to have been established.

Assuming post-shock bulk energy redistribution takes place through Coulomb collisions,
two stages are expected,
first amongst ions and then between ions and electrons.
During both these stages, the ionization state will also
be moving from that of the pre-shock gas towards its relaxed state.
Post-shock equilibrium would apply once all
these various stages have run to completion. We suggest
that the time-scales are too long for this to happen.

\cite{S:1962}, \citet[][section 6.4]{KT:1986} and \citet{ZR:2002}, for example, showed that
equilibration between ions and electrons is much slower than between electrons or ions among themselves
because of the large difference in mass between ions and electrons.
\cite{S:1962} estimated about a factor of 50 slower than ion-ion equilibration
from the square root of the mass ratio
for modest departures from equipartition.

Immediately behind a stellar wind shock, conditions are far from equilibrium.
\citet{BPS:2005} have made rigorous calculations of the
rate of temperature equilibration, incorporating
simultaneous Coulomb scattering on short scales and collective plasma
heating on large scales, 
for arbitrary ion and electron temperatures, $T_{i}$ and $T_{e}$.
Energy transfer is governed by the term
\begin{equation}\label{Equation:IonElectronEquilibration}
d(T_{i}-T_{e})/dt \propto \kappa_i^2\kappa_e^2\frac{(\beta_{i}m_{i}\beta_{e}m_{e})^{1/2}}{(\beta_{i}m_{i}+\beta_{e}m_{e})^{3/2}}
\end{equation}
where $\kappa_{i}^2=\beta_{i}e_{i}^2n_{i}$ and $\beta_{i}=1/T_{i}$.
This equation shows that in the very earliest stages of relaxation, when
ion and electron random velocities are similar and \hbox{$T_{i} \gg T_{e}$},
the electrons will initially heat up quickly in a time proportional to $(m_{i}m_{e})$
intermediate between those of electrons and ions among themselves.
As the electrons heat up, the fraction of energy $(m_{e}/m_{i})$ transfered
in a collision has less effect and the rate slows down to the factor
$\sqrt{m_{e}/m_{i}}$ times ion-ion rate calculated by Spitzer.
    
Thus, the length scale for post-shock electron heating
in the wind of an O star compared to the ion-ion mean-free-path of Fig~\ref{Figure:ion-ion:mean-free-path}
is likely to be many stellar radii
in which case the hot gas may not be able to survive long enough for
equilibrium to be established.
The low X-ray
emission measures of O-stars
show that only a small fraction of the wind
material is involved, the vast bulk of which is cold:
shocked gas will soon get mixed back with cool material
and disappear from view. In this time, there will be little
chance for electrons to be heated enough to be able to
contribute to the observed X-ray emission by either ionization or
bremsstrahlung continuum. A scarcity of
shock-heated electrons in the winds of O stars would
give rise to a weak continuum and require an alternative means
of producing the high levels of ionization.

\subsection{X-ray ionization by ion-ion interactions in O-star winds}

In the supposed absence of shock-heated electrons, what is the
origin of the ion species unambiguously identified in the X-ray spectrum?
Ions in general and protons in particular
in the immediate post-shock gas are probably responsible,
as anticipated in general terms by \citet[][p. 148]{M:1999}.
Through the initial shock transition, the ionization balance is
not changed although ions characteristic of the partially
ionized cool wind
suddenly find themselves in a hostile environment:
encounters take place with other ions at relative velocities
caused by randomization of the bulk velocity of the pre-shock gas
of about $2000~{\rm km}~{\rm s}^{-1}$ in the
terminal velocity regime of \hbox{$\zeta$~Orionis}. Protons of such
velocities are a similarly effective agent for ionization and excitation
as electrons of a few keV because the cross-sections depend on the
relative velocity of the incident ionizing particle and the
velocity of the bound electron,
whose order of
magnitude is fixed by the Bohr velocity
\hbox{$v_{\rm Bohr}=2188~{\rm km}~{\rm s}^{-1}$}.
We suggest that it is the
coincidence of this microscopic atomic value of $v_{\rm Bohr}$
- widely used as the atomic unit of velocity -
and the macroscopic
terminal velocities of \hbox{$\zeta$~Orionis} and other O stars,
that is the basic physical reason for the production of X-rays in
hot stars.
 
The relaxing material contains no neutral atoms but is still far from the
fully-ionized state discussed by Spitzer and others, so that the same
ion-ion interactions involved in energy exchange via Coulomb collisions
are likely simultaneously to affect the remaining bound electrons, with
possible observational consequences.
The ion-ion Coulomb cross-section of $1/n_{\rm i} l_{\rm i-i} \sim 7\times10^{-21}~{\rm cm}^2$ estimated
from equation~(\ref{Equation:lii}),
which determines the rate of energy
exchange in elastic collisions between ions in the post-shock gas, serves as a benchmark
in the discussion below against which the effectiveness may be judged of
inelastic processes that also occur during relaxation.
High-velocity encounters between ions have maximum inelastic cross-sections of order
$10^{-16}~{\rm cm}^2$
that exceed the Coulomb cross-section by orders-of-magnitude
and thus might be expected to dominate post-shock relaxation as discussed in more quantitative terms below.

Ion-ion interactions that involve bound electrons
are of three types, namely
ionization, excitation and charge exchange. Ionization and charge exchange are both
mechanisms of electron loss that lead to changes in the ionization
state and so could be responsible for the highly-charged ions
observed in the X-ray spectrum. Excitation and sometimes charge exchange
leave ions in excited states from which decay leads to observable radiation including X-rays.
A typical recombination cross-section
is of order $10^{-21}~{\rm cm}^2$ and too small to compete
with radiative de-excitation in X-ray production.

\subsubsection{Ion-ion ionization}

Ion-atom ionization has been discussed in detail by \cite{KSD:2006},
whose scaling law for the ionization cross-section by a fully-stripped
projectile ion of charge $Z_p$ and velocity $v$ of one of the
$N_{nl}$ electrons bound in a quantum
state $(n,l)$ of ionization potential $I_{nl}$
to a target nucleus of charge $Z_T$ is given by
\begin{equation}\label{Equation:IonAtomIonization}
\sigma^{ion}(v,I_{nl},Z_p)={\pi}a_0^2\frac{Z_p^2}{(Z_p/Z_T+1)}N_{nl}\frac{E_0^2}{I_{nl}^2}G\left(\frac{v}{v_{nl}\sqrt{Z_p/Z_T+1}}\right)
\end{equation}
where atomic units for cross-section, velocity and energy are defined by the Bohr cross-section
${\pi}a_0^2=0.880\times10^{-16}~{\rm cm}^2$;
$v_0=v_{\rm Bohr}=2188~{\rm km/s}$; and
$E_0=m_{\rm e}v_0^2=27.2~{\rm eV}$, respectively.
This equation shows in particular how ion-ion ionization cross-sections
depend sensitively through the scaling function $G(v/v_{\rm max})$ on the relative velocity of the incident ionizing particle
and the bound electron. The maximum cross-section, $\sigma_{\rm max}$,
is given by $G(1)\sim0.8$ at $v = v_{\rm max} = v_{nl}\sqrt{Z_p/Z_T+1}$.

The velocity of a bound electron is roughly $v_{nl}/v_0 \sim Z_T/n \sim \sqrt{2I_{nl}/E_0}$.
In the immediate post-shock gas of a stellar wind, the relative velocity scale of ion-ion encounters
is fixed by the randomized terminal velocity $v_{\infty}$, though the Maxwellian distributions will ensure that
a fraction of encounters also take place at relative velocities up to a few times this value.
\cite{KSD:2006} showed that their scaling law
\begin{equation}\label{Equation:IonizationVelocityLaw}
G(x)=\frac{\exp{(-1/x^2)}}{x^2}[1.26+0.283\ln{(2x^2+25)}],
\end{equation}
which falls away from the maximum towards the lower velocities of interest here,
reproduces well a variety of observational data for $x > 0.5$,
at which point the ionization cross-section has fallen to 20\% of its maximum value.
At low velocities, measurements become
difficult because, to quote Kaganovich et al., the ionization cross-section is completely
dominated by charge exchange, whose cross-section is comparable to $\sigma_{\rm max}$.

\subsubsection{Ion-ion charge exchange}

Charge exchange between ions is likely to be of fundamental importance in the immediate post-shock gas
in the wind of a hot star and be responsible for production of many highly-charged ions.
Following \citet[][pp. 11-15]{BM:1992}, a simple theoretical estimate for
the low-velocity charge exchange cross-section
between a bare nucleus projectile of charge $Z_p$ interacting with a target nucleus of charge $Z_T$ with
a single bound electron of principal quantum number $n$ may be written
\begin{equation}\label{Equation:ChargeExchange}
\sigma^{CE}(Z_p,Z_T)={\pi}a_0^2\left(\frac{2n^2(Z_T+2\sqrt{Z_pZ_T})}{Z_T^2}\right)^2
\end{equation}
independent of velocity for $v/v_{nl} < (Z_p/4Z_T)^{\frac{1}{4}}$.
This expression reproduces experimental data well for electron capture by ions from
atomic hydrogen and if applied to collisions between protons ($Z_p=1$) and, say,
${\rm O}^{7+}$ ($Z_T=8$) ions, predicts a cross-section of $1.6\times10^{-17}{\rm cm}^{-2}$.
Crossed-beam measurements of collisions between other ions have been reported, for example, by
\citet{STH:etal:2001}, \citet{vD:etal:2001} and \citet{B:etal:2005} and show similar or higher cross-sections.
including more than $10^{-16}{\rm cm}^{-2}$ for
the production of ${\rm N}^{5+}$ from charge exchange with \hbox{$\alpha$-particles}:
this is interesting because \ion{N}{vi} lines appear in the X-ray spectrum of \hbox{$\zeta$~Orionis}.

\subsubsection{Implications for post-shock relaxation and the X-ray spectrum}

These high cross-sections, which overwhelm Coulomb collisions, guarantee that the first thing
to happen to ions in the post-shock relaxation layer
concerns their outer bound electrons with
charge exchange and ionization leading to rapid
readjustment of the ionization balance.
This even takes place faster
than relaxation among different ions to a common temperature.
This conclusion is probably of more general significance.
In SNRs, as calculated by \citet{HH:1985} for example, it has often been
assumed that the post-shock ionization balance changes
in response to hot electrons, whereas particularly for the shocks velocities
of a several thousand ${\rm km}~{\rm s}^{-1}$ of some of the remnants
discussed by \citet{R:2006}, ion-ion interactions are also important,
as in the charge-exchange layer immediately behind some SNR shocks
discussed by \citet[][p. 403]{DM:1993}, for example.

For O-star winds, we propose that it is also ion-ion interactions and notably charge exchange
during the initial phase of post-shock
relaxation that give the observed X-rays.
While there seems little doubt that the necessary ions can be produced,
the generation of the X-ray spectrum is more uncertain. This may be due to
excitation by ions, mainly protons; charge exchange into excited states in the hot gas;
or charge exchange in eventual encounters with the much less tightly
bound electrons in the plentiful supply of much cooler ions in nearby unshocked material
that makes up the great majority of the wind.
This final possibility seems worthy of future work.
If proton ionization is significant,
while it is not expected that the great majority of
free electrons will be heated much immediately behind the shock,
a small population of
hot secondary electrons would be ejected,
giving some scope for a bremsstrahlung continuum.

\subsection{The velocity profile of the X-ray lines}

Most of the efforts made so far to account for the shape of the
X-ray lines in hot stars, such as those of \cite{IG:2002},
\cite{KCO:2003} and \cite{OFH:2004}, have assumed that the X-ray
emitting gas is moving with the majority cool gas. By doing so and
appealing to the velocity law of the cool gas, attempts have been
made to constrain the location of the hot gas. In particular, it has
regularly been argued that the red wing of the X-ray lines arises in
material in the wind on the far side of the star flowing away from
the observer. This simple assumption is hard to justify, if only
because much of the ordered motion of the wind needs to be converted
to heat for X-rays to be observed at all and, for CIE models,
half the energy has to be transfered from ions to electrons for excitation to occur.

For an ideal gas of mean
particle mass $m$ flowing at velocity $v$ into a strong stationary
shock, the jump conditions determine the equilibrium temperature and
velocity of the outflowing shocked gas to be
$kT_{\rm S}/m=(3/16)v^2$ and $v_{\rm S}=v/4$. For material of solar
composition, $kT_{\rm S}\sim1.2v_8^2~{\rm keV}$, where
$v=v_8\times1000~{\rm km}~{s}^{-1}$, so that the observation of X-ray
gas in equilibrium near the \hbox{0.5~keV} of Table~\ref{Table:TriLine:model}
would imply the complete dissipation of
about $650~{\rm km}~{\rm s}^{-1}$ of directed kinetic energy.

Out of equilibrium,
it is more likely that the observed line velocity profiles mostly
reflect instead the line-of-sight component of the thermalized
motion of ions in the immediate post-shock gas. For a Maxwellian
distribution,
\begin{equation}\label{Equation:HWHM}
{\rm HWHM}(v_{\rm x})=\sqrt{2\ln{2}(kT_{\rm S}/m)} = (\sqrt{6\ln{2}}/4)v \sim 0.51v.
\end{equation}
The similar observed widths of different ions further implies
that even equilibration between ions does not take place.

In addition to this thermal component, the bulk flow away from the
shock at $v_{\rm S}=v/4$ will also contribute to the line profile. The
calculations that \cite{IG:2002} made for material moving at
constant radial velocity in the terminal velocity flow, may equally
well be applied to this slower-moving material. The resultant,
roughly triangular, line profiles show blue-shifted peaks at 10 or
20\% of the outflow velocity, depending on the optical depth, and a
half width of about 60\%. The resultant combination of thermal and
bulk motions should then shift the line to the blue by a
fraction of the terminal velocity with a width mostly determined by
the greater random motion.

The observed lines in \hbox{$\zeta$~Orionis} are roughly consistent
with this scheme, showing blue-shifted peaks of about $v_{\infty}/7$ and
half-widths of about $80\%$ of the value of $0.51v_{\infty}$
expected for complete thermalization of gas moving at the terminal
velocity. 
While it is conceivable that the X-ray emission is picking
out the small amount of gas in the wind-acceleration zone travelling
at this fraction of the terminal velocity, we consider this
unlikely, not least because of the apparent lack of X-ray absorption..
Two further possibilities
come to mind: either that the shocks are not stationary in the wind
but are moving outwards at a few hundred \hbox{${\rm km}~{\rm
s}^{-1}$} and are completely thermalized; or that the shocks are
stationary and occur throughout the
terminal velocity regime and the energy balance involves other
channels than merely the production of heat. As non-thermal particle
acceleration is a common feature of collisionless shocks, such as
those in SNRs and those probably in WR~140, we consider this
possibility the more promising, although free-free emission and
absorption by the cool wind is expected to be strong enough to
prevent any non-thermal radio radiation from being observed.

\subsection{Absence of equilibrium in the post-shock plasma and the He-like triplets}

If O-star X-rays are produced by terminal-velocity shocks in the wind, then the
sluggishness of energy exchange throughout the post-shock gas will
prevent the establishment of equilibrium before the hot
gas is overwhelmed by cool material. Therefore, the use of coronal
equilibrium models to analyse O-star X-ray spectra is unlikely to
yield any physically meaningful information about conditions in the
hot gas, though it is always possible to derive equivalent
equilibrium temperatures like those quoted above for
\hbox{$\zeta$~Orionis} and elsewhere for other stars. The X-ray
spectrum is more likely to reflect the individual cross-sections of
the three important processes between ions discussed above of
charge exchange, ionization and excitation.

An argument in support of the instability-driven shock
model has been its success following \cite{KLC:etal:2001} and \cite{WC:2001}
in accounting
for the characteristic patterns of the {\it fir} lines of the He-like
triplets: in contrast to most other high-resolution X-ray spectra,
the O-star intercombination lines are particularly strong, showing
the effects of photospheric UV pumping of the $^3{\rm S}_1$
upper level of the forbidden line to $^3{\rm P}_{1,2}$ upper levels
of the intercombination line.
In \hbox{$\zeta$~Puppis}, \cite{KLC:etal:2001} showed that there is
enough radiation that the photoexcitation rate exceeds the
intercombination line decay rate with a radius that varies from about
$3R_*$ for \ion{Si}{xiii} to $200R_*$ for \ion{N}{vi},
so that the UV affects an enormous volume of the wind.

Calculations of the expected line ratios such a UV field, such as those of \cite{PMDRK:2001},
have been based on otherwise equilibrium conditions.
In the rarefied gas immediately behind the collisionless shock in an
O-star wind imagined above, apart from the intense UV radiation environment,
other conditions are expected to be quite
different with the He-like ratios subject to a variety of considerations, namely:
\begin{itemize}
   \item a plasma out of equilibrium
   \item a scarcity of hot electrons
   \item energy concentrated in interacting ions
   \item widespread charge exchange
   \item ionization by non-thermal particles
\end{itemize}
Hot protons and cool electrons are expected to have
similar random velocity distributions and thus both contribute
to collisional coupling of the He-like triplet levels.
Proton excitation leads to different line ratios
because of the absence of resonances and exchange terms \citep[e.g.][p. 41]{SV:1993}.

Charge exchange is likely to be as important in stellar winds as it is
for the
generation of cometary X-rays by charge exchange of neutral material
with highly-charged ions in the solar wind.
In laboratory measurements designed to simulate cometry charge exchange by \cite{BBB:etal:2003},
which are at very low velocities on the atomic scale,
the forbidden line was strongly enhanced.
In the crossed-beam measurements discussed above at higher relative velocities more relevant to hot stars,
though still low on the atomic scale,
the total cross-sections are dominated by charge exchange.

Although, following the suggestion by
\cite{P:1987}, most authors have cited inverse-Compton emission as
a non-thermal signature, ionization losses are
a more important cooling mechanism of mildly relativistic particles.
This may be an opportunity to observe, at least
indirectly, the effects of cosmic-ray acceleration in
single stellar winds through the detection of
satellite lines following inner-shell
ionization \citep[e.g.][]{MS:1978},
which may also cause the unusually strong He-like forbidden lines
in the colliding-wind spectra $\eta$~Carinae and WR~140 \citep{PCSW:2005}.
Incidentally, there may be some evidence of satellite lines
in the \ion{O}{vii} triplet of \hbox{$\zeta$~Orionis} in
which the intercombination line appears
significantly blue-shifted from the wavelength expected from the
ensemble of other lines by \hbox{$-13\pm3$~{\rm m}\AA}.
This partly accounts for the extra blue shift of \ion{O}{vii}
in Table~\ref{Table:TriLine:ions}.

In summary, some theoretical effort will be required to
calculate the X-ray spectrum in general and the He-like line ratios in particular 
to be expected from shocked ions in an O-star wind
before proper conclusions can be drawn about the effects of UV
radiation and the implications for the location of the X-ray emitting material.

\section{X-ray absorption in O-star winds}

In spite of the high optical depths expected on the basis of
supposedly well-determined mass-loss rates, \cite{KLC:etal:2001}
showed the lack of any detectable absorption edges in
\hbox{$\zeta$~Puppis}. Even cursory inspection of
\hbox{$\zeta$~Orionis}'s RGS spectrum or those of other O-stars
\citep[e.g.][]{PK:2003} shows no obvious sign of absorption, as
emphasized by the strength of \ion{C}{vi} Ly$\alpha$. 

An important part of
attempts to reconcile the observational data with an origin deep in
the wind's acceleration zone within a few stellar radii of the
surface has involved
seeking means of reducing the X-ray opacity.
From a theoretical point-of-view,
\cite{WC:2001} and \cite{CMWMC:2001} used the reduced opacities
of \cite{WCDS:1998} from self-consistent calculations
of the ionization balance to be expected in the presence
of UV and X-radiation \citep{M:etal:1993,MCW:1994}.
Independently, \cite{KCO:2003}, for example,
appealed to significantly reduced attenuation in their
models of the \hbox{$\zeta$~Puppis} X-ray line profiles.

It might have been expected that the high UV outer-shell optical
depths necessary for the line driving would guarantee a
correspondingly high X-ray inner-shell optical depth. It has been
known for many years that this is indeed the case,
as demonstrated by the bright neutron star hard
X-ray sources powered by accretion from an otherwise normal hot
star. Vela X-1 (e.g. \cite{SLKP:1999} and references cited there)
is a relevant example as the primary star HD~77581 is a B0.5Iab star
of similar spectral type to \hbox{$\zeta$~Orionis}. The neutron star
orbits at about $0.78R_*$ above the surface of the supergiant, in
what would otherwise be the strongest region of the acceleration
zone. It is a simple observational fact that
the neutron-star
secondary's X-rays are subject to heavy absorption.
While this varies through the orbit, reaching values of several times $10^{23}{\rm cm}^{-2}$,
much of it is due to the rearrangement of wind material into an accretion wake
that trails the neutron star.
The X-ray absorption, though occasionally extremely high, is well modelled
by cold material.

Immediately after eclipse, on the other hand,
when the line-of-sight passes only through the stellar wind,
a minimum column density of about $2\times10^{22}{\rm cm}^{-2}$
is observed..
This value is reasonably close to that expected if \hbox{$\zeta$~Orionis} were to take the place of HD~77581.
For a spherically symmetric wind with a \hbox{$\beta$ velocity} law, the integrated column density
between a point distance $R$ from the surface of the star along the line-of-sight scales as
\begin{equation}
N_R(R,\phi,i) \sim (\dot{M}/\mu m_p v_{\infty} R_*) (\gamma/\sin{\gamma}) H(R,R_*,\beta) \ {\rm cm}^{-2}
\end{equation}
where $\mu$ is the mean atomic number, $m_p$ the proton mass and
\hbox{$v(r)=v_{\infty}(1-R_*/r)^{\beta}$}.
The column density is the product of a column density characteristic
of the star, $N_* = (\dot{M}/\mu m_p v_{\infty} R_*)$,
and two scaling factors:
a geometrical factor $(\gamma/\sin{\gamma})$,
where \hbox{$\cos{\gamma}=\cos{\phi}\sin{\theta}$} and 
$\phi,\theta$ are the azimuth and inclination angles between the radial direction through $R$ and the line-of-sight;
and an analytical density-related factor
$H(R,R_*,\beta)$, which exceeds unity for points in the densest parts of the wind and equals $(R_*/R)$ far from the stellar surface.
For the parameters of \hbox{$\zeta$~Orionis} in Table~\ref{Table:zO:data},
$N_* \sim 7.3\times10^{21}{\rm cm}^{-2}$.
For the line-of-sight that would apply to a
neutron star emerging from an eclipse by \hbox{$\zeta$~Orionis}, $\gamma\sim\pi/2$
leading to $N_R\sim1.1\times10^{22}{\rm cm}^{-2}$, only about a factor of two lower the observed value in Vela X-1.
The evidence, therefore, is that X-rays produced at $R=1.78R_*$ in the
wind of a supergiant companion of similar type to \hbox{$\zeta$~Orionis}
are absorbed. This would equally be expected to apply
to any intrinsic X-rays produced here and lead to the rapidly increasing
emission measures and decreasing velocities plotted in Fig.~\ref{Figure:LineProperties}.

\section{A new paradigm for the X-ray emission from O stars}

We propose that the high inertia of the plasma flow in O-star winds, caused
by the weakness of Coulomb interactions, has important consequences for
the production of X-rays in the flow. If shocks develop at all, they
are collisionless shocks in which the fundamental role of plasma
instabilities suggests that the local magnetic field could be the
``missing variable'' responsible for the scatter of an order-of-magnitude or more in the modest X-ray luminosities
generated by otherwise apparently similar O stars.

The long Coulomb collisional mean-free-path, apart from requiring a collisionless
shock transition and slow post-shock energy exchange, also more generally
limits the steepness of pressure gradients that can be sustained by the medium
and defines a minimum size for any structures in the hot gas. In this case,
it is unlikely that the microscopic instability in the line-driving mechanism
will be able to steepen into the macroscopic shocks widely thought to
generate O-star X-rays: the instability would soon be limited by the
difficulty of transferring momentum quickly enough from the unstable driven ions to the rest
of the wind, causing breakdown of the single-fluid approximation as discussed, for example,
for thin hot-star winds by \cite{SP:1992}. Furthermore, the narrowness of the
X-ray line profiles would appear to exclude the type of positive-going velocity
instabilities in the calculations of \cite{OCR:1988}, for example.

The observed common line profile implies a common seat of X-ray emission for all
the ions, a notion qualitatively consistent with an origin in the terminal velocity regime.
While X-rays are probably produced throughout the wind as a routine aspect of the chaotic, supersonic
flow of magnetized plasma,
the observable volume is likely to be limited to some extent
by photoelectric absorption,
giving rise to small systematic changes in the line profiles from ion to ion.
The line width is mainly due to randomization of the directed kinetic
energy of the flow.

A thorough quantitative assessment of these ideas will have to await a
detailed emission model that is expected to include, among other things,
ionization, excitation and charge exchange averaged over the microscopic
velocity distribution of ions in the post-shock gas as well as an energy budget with
a non-thermal component.

An inevitable consequence of the scheme proposed is the effective absence of hot electrons
in an O-star wind, altering the physical basis of the plasma emission models
concerned. In contrast to either ionization by electron impact or photoionization,
which together account for the majority of observed X-ray plasmas, single O-star spectra
may be one of the clearest examples of a protoionized plasma.
The term ``protoionized'' seems appropriate both for the contrast
with photoionized and because it describes the very earliest
stages of post-shock relaxation through which tenuous plasmas are bound to pass
before electrons become hot enough to take over.
The shock transitions
themselves, though probably smaller in physical extent than the extensive shocks which span
a large-scale colliding-wind flow such as WR~140, initially obey similar jump conditions.
The distinction between spectra lies rather in the post-shock layer, in the amount of equilibration
that takes place between ions and electrons. If the hot plasma is confined by magnetic fields, as
in WR~140, for example, post-shock relaxation may take its long-term
course allowing energy to be transfered from ions
to electrons, which may then excite a familiar
plasma spectrum characteristic of collisional ionization equilibrium.
Otherwise, in the winds of single O-stars,
the plasma is not in equilibrium,
few or no electrons reach high temperatures
and we observe instead the effects of
the coincidence between the macroscopic terminal velocity of the wind
and the microscopic Bohr velocity characteristic of electrons in bound atomic states.
Interactions with the majority cool material may also be important.

There are two simple tests of the new scheme which the data available thus far are not quite able to support:
long, high-resolution, exposures of single O-stars should be able to establish two characteristics of the X-ray
spectrum: first, that all the lines have much the same shape; and second that the continuum
is weaker than that from an electron-excited plasma.

\begin{acknowledgements}
Many thanks are due to Ton Raassen of SRON Utrecht in the Netherlands, who cast a critical eye over many stages of
the development of the ideas discussed here.
\end{acknowledgements}

\clearpage
\onecolumn

\begin{center}
\topcaption{\label{Table:Line:fluxes}
            Fluxes of the X-ray lines in \hbox{$\zeta$~Orionis} using the {\tt TriLine} model of Table~\ref{Table:TriLine:model} for each line.}
\tablefirsthead{
   \hline
   \multicolumn{1}{l}{Ion}  & \multicolumn{1}{c}{$\lambda_{\rm lab}$}
                            & \multicolumn{1}{c}{Line ID}
                            & \multicolumn{2}{c}{Flux($10^{-5}{\rm cm}^{-2}{\rm s}^{-1}$)} \\
   \hline}
\tablehead{
   \hline
   \multicolumn{1}{l}{Ion}  & \multicolumn{1}{c}{$\lambda_{\rm lab}$}
                            & \multicolumn{1}{c}{Line ID}
                            & \multicolumn{2}{c}{Flux($10^{-5}{\rm cm}^{-2}{\rm s}^{-1}$)} \\
   \hline}
\tabletail{\hline}
\tablelasttail{\hline}
\begin{supertabular}{lrlr@{$\pm$}l}
\ion{C}{vi}     & 33.734 & Ly$\alpha$   & 40.90  & 1.72     \\
                & 28.465 & Ly$\beta$    &  6.76  & 0.73     \\
                & 26.990 & Ly$\gamma$   &  1.59  & 0.49     \\
\hline
\ion{S}{xii}    & 36.398 &              &  4.66  & 0.97     \\
                & 36.573 &              &  1.31  & 0.88     \\
\hline
\ion{S}{xiii}   & 35.667 &              &  6.96  & 0.94     \\
                & 37.598 &              &  3.99  & 1.78     \\
                & 32.242 &              &  1.93  & 0.68     \\
\hline
\ion{N}{vi}     & 28.787 &  {\it r}     &  7.88  & 0.79     \\
                & 29.084 &  {\it i}     &  7.48  & 0.85     \\
                & 29.534 &  {\it f}     &  0.00  & 0.25     \\
                & 24.898 &  He$\beta$   &  1.35  & 0.73     \\
\hline
\ion{N}{vii}    & 24.779 & Ly$\alpha$   & 17.64  & 0.72     \\
                & 20.909 & Ly$\beta$    &  0.26  & 0.47     \\
\hline
\ion{O}{vii}    & 21.602 &  {\it r}     & 96.13  & 2.73     \\
                & 21.807 &  {\it i}     & 75.99  & 2.47     \\
                & 22.101 &  {\it f}     &  7.69  & 0.96     \\
                & 18.627 &  He$\beta$   &  8.41  & 0.57     \\
                & 17.768 &  He$\gamma$  &  1.98  & 0.36     \\
                & 17.396 &  He$\delta$  &  1.60  & 0.36     \\
\hline
\ion{O}{viii}   & 18.967 &  Ly$\alpha$  & 92.43  & 1.73     \\
                & 16.005 &  Ly$\beta$   & 13.55  & 0.75     \\
                & 15.176 &  Ly$\gamma$  &  7.05  & 0.75     \\
                & 14.820 &  Ly$\delta$  &  1.12  & 0.29     \\
\hline
\ion{Fe}{xvii}  & 15.015 &              & 37.68  & 1.10     \\
                & 15.262 &              & 17.36  & 0.83     \\
                & 15.453 &              &  4.45  & 0.45     \\
                & 16.778 &              & 19.45  & 0.70     \\
                & 17.053 &              & 30.05  & 1.65     \\
                & 17.098 &              & 17.03  & 1.65     \\
                & 12.266 &              &  1.92  & 0.69     \\
\hline
\ion{Fe}{xviii} & 14.208 &              &  7.77  & 0.42     \\
                & 14.373 &              &  4.03  & 0.36     \\
                & 14.534 &              &  2.79  & 0.33     \\
                & 15.625 &              &  2.91  & 0.39     \\
                & 15.870 &              &  1.93  & 0.35     \\
                & 16.071 &              &  2.96  & 0.61     \\
                & 17.623 &              &  2.76  & 0.38     \\
\hline
\ion{Fe}{xix}   & 13.518 &              &  2.54  & 1.26     \\
                & 13.795 &              &  3.99  & 0.38     \\
                & 15.079 &              &  1.23  & 0.90     \\
\hline
\ion{Ne}{ix}    & 13.447 &  {\it r}     & 20.35  & 0.82     \\
                & 13.553 &  {\it i}     & 11.95  & 1.07     \\
                & 13.699 &  {\it f}     &  3.14  & 0.39     \\
                & 11.547 &  He$\beta$   &  2.22  & 0.22     \\
                & 11.000 &  He$\gamma$  &  0.97  & 0.17     \\
\hline
\ion{Fe}{xx}    & 12.824 &              &  0.45  & 0.08     \\
                & 12.846 &              &  0.45  & 0.08     \\
                & 12.864 &              &  0.45  & 0.08     \\
\hline
\ion{Fe}{xxi}   & 12.284 &              &  1.27  & 0.29     \\
\hline
\ion{Ne}{x}     & 12.132 &  Ly$\alpha$  & 12.02  & 0.45     \\
                & 10.238 &  Ly$\beta$   &  1.35  & 0.16     \\
                &  9.708 &  Ly$\gamma$  &  0.33  & 0.16     \\
                &  9.481 &  Ly$\delta$  &  0.07  & 0.15     \\
\hline
\ion{Mg}{xi}    &  9.169 &  {\it r}     &  2.86  & 0.18     \\
                &  9.231 &  {\it i}     &  1.31  & 0.15     \\
                &  9.314 &  {\it f}     &  1.25  & 0.13     \\
                &  7.851 &  He$\beta$   &  0.26  & 0.12     \\
\hline
\ion{Mg}{xii}   &  8.419 &  Ly$\alpha$  &  0.55  & 0.13     \\
                &  7.106 &  Ly$\beta$   &  0.10  & 0.09     \\
\hline
\ion{Si}{xiii}  &  6.648 &  {\it r}     &  0.98  & 0.12     \\
                &  6.688 &  {\it i}     &  0.38  & 0.14     \\
                &  6.740 &  {\it f}     &  0.89  & 0.11     \\
\hline
\ion{Si}{xiv}   &  6.180 &  Ly$\alpha$  &  0.27  & 0.11     \\
\hline
\end{supertabular}
\end{center}

\end{document}